\documentclass[journal]{IEEEtran}

\ifCLASSINFOpdf
   \usepackage[pdftex]{graphicx}

\else
   \usepackage[dvips]{graphicx}
\fi

\usepackage{cite}

\usepackage[draft,markup=default]{changes}

\usepackage{amsmath}

\ifCLASSOPTIONcompsoc
  \usepackage[caption=false,font=normalsize,labelfont=sf,textfont=sf]{subfig}
\else
  \usepackage[caption=false,font=footnotesize]{subfig}
\fi

\begin{document}

\title{Coupling Matrix Representation of Nonreciprocal \\
       Filters Based on Time Modulated Resonators}

\author{Alejandro~Alvarez-Melcon,~\IEEEmembership{Senior~Member,~IEEE,}
    Xiaohu~Wu,~\IEEEmembership{Member,~IEEE,}
        Jiawei~Zang, Xiaoguang~Liu,~\IEEEmembership{Senior~Member,~IEEE,}
        and~J.~Sebastian~Gomez-Diaz,~\IEEEmembership{Senior~Member,~IEEE}
\thanks{Manuscript received Month DD, YYYY; revised Month DD, YYYY;
accepted Month DD, YYYY.}
\thanks{This work is supported in part by the National Science Foundation
with CAREER Grant No. ECCS-1749177 and by the
grants PRX18/00092 and TEC2016-75934-C4-4-R of MECD, Spain. }
\thanks{A. A. Melcon was on sabbatical in the Department of Electrical
and Computer Engineering at UC Davis. He is now with the Department
of Information and Communication Technologies, Technical University
Cartagena, 30202, Spain. e-mail: alejandro.alvarez@upct.es.}
\thanks{X. Wu is with the
Department of
Electrical and Computer Engineering, University of
California, Davis, USA. e-mail: xiaohu.wu@nuist.edu.cn, wxhwu@ucdavis.edu}
\thanks{J. Zang was a visiting student in the Department of
Electrical and Computer Engineering at UC Davis. He is now with
the School of Information and Electronics,
Beijing Institute of Technology, Beijing 100081, China.}
\thanks{J. S. Gomez-Diaz and X. Liu are with the Department of
Electrical and Computer Engineering, University of California,
Davis, USA. e-mail: \{jsgomez, lxgliu\}@ucdavis.edu}}

\markboth{Journal of \LaTeX\ Class Files,~Vol.~xx, No.~x, August~20xx}%
{Time Modulated Nonreciprocal Filters}

\maketitle

\begin{abstract}
This paper addresses the analysis and design of non-reciprocal filters
based on time modulated resonators. We analytically show that time
modulating a resonator leads to a set of harmonic resonators composed
of the unmodulated lumped elements plus a frequency invariant
element that accounts for differences in the resonant frequencies.
We then demonstrate that harmonic resonators of different order
are coupled through non-reciprocal admittance inverters whereas
harmonic resonators of the same order couple with the admittance
inverter coming from the unmodulated filter network. This coupling
topology provides useful insights to understand and quickly design
non-reciprocal filters and permits their characterization using
an asynchronously tuned coupled resonators network together with the
coupling matrix formalism. Two designed filters, of orders three
and four, are experimentally demonstrated using quarter wavelength
resonators implemented in microstrip technology and terminated
by a varactor on one side. The varactors are biased using
coplanar waveguides integrated in the ground plane of the device.
Measured results are found to be in good agreement with numerical
results, validating the proposed theory.
\end{abstract}

\begin{IEEEkeywords}
Coupling matrix, microwave filters, non reciprocity,
spatio-temporal modulation, time modulated capacitors.
\end{IEEEkeywords}

\IEEEpeerreviewmaketitle

\section{Introduction}

\IEEEPARstart{N}{on-reciprocal} components are of key importance
in many electronic systems, such as radar or mobile and satellite
communications \cite{pozarlibro}. Traditionally, such components
have relied on magnetic materials, such as ferrites, under strong
biasing fields. Increasingly stringent technological demands, in
constant pursuit of integration, affordability, and miniaturization,
have triggered the recent emergence of magnetless non-reciprocal
approaches to break the Lorentz reciprocity principle \cite{caloz18}
and the subsequent development of devices such as circulators
\cite{kord18,kodera13,romain14,estep14,estep16,reiskarimian16,reiskarimian17,dinc17,ahmed18b,ahmed18c,yu18},
isolators \cite{yu08,lira12,qin14,chang15,biedka17,correas18,sounas18},
and even non-reciprocal leaky-wave antennas
\cite{correas16,yakir16,taravati17} operating in the absence
of magneto-optical effects.

In this context, the concept of non-reciprocal filters based on
time-modulated resonators have recently been put forward \cite{wu18}.
The operation principle
behind this type of filters relies on tailoring the non-reciprocal
power transfer among the RF and intermodulation frequencies to create
constructive/destructive interferences at the input/output ports.
The filters were analyzed in \cite{wu18} through a dedicated spectral
domain method combined with ($ABCD$) parameters,
and useful guidelines on how to optimize the frequency,
amplitude, and phase delay of the signals that modulate the resonators
were given. A practical prototype was also experimentally
demonstrated using varactors and lumped inductors.

{\color{black}
Here we should remark that the combination of time modulated resonators with
sinusoidal modulation signals will enhance the generation of the
two first intermodulation products (the so called $+1$ and $-1$
harmonics \cite{qin14}). The minimization of higher order
intermodulation products may be interesting, since it will help
to keep under control the power conversion between harmonics,
and to simplify the
tailoring process needed to create constructive/destructive
interferences at the input/output ports.
}

In this paper, we develop a coupling matrix representation of non-reciprocal
filters based on time modulated resonators. Starting from the initial
unmodulated equivalent circuit, a multi-harmonic equivalent network
is rigorously derived, taking into account the nonlinear harmonics
(also known as intermodulation products) that are internally excited.
By introducing the concept of harmonic resonators, the resulting
structure is represented with a simple network based on a specific
coupled resonator topology. It is analytically shown that resonators
of identical harmonic orders are coupled with the admittance
inverters found in the original unmodulated network while
resonators of different harmonic orders are coupled through
non-reciprocal admittance inverters. In addition, analytic formulas
are derived to represent the new harmonic resonators with
{\color{black} Frequency
Invariant Susceptances}  \cite{cameron03,cameronlibrogeneric} that
accounts for differences in the resonant frequencies. In this way,
all the resonators of the resulting network are expressed
in terms of the original unmodulated resonators.

It is important to emphasize that the analytic calculation of the
non-reciprocal admittance inverters and the
{\color{black} frequency
invariant susceptances} for
harmonic resonators, together with the derived coupling topology,
permits to analyze and design non-reciprocal filters using an
asynchronously tuned coupled resonator network and the classical
coupling matrix formalism \cite{cameronlibrogeneric}.
{\color{black}
Here the term {\em asynchronously tuned} is used to refer
to coupling
topologies having resonators tuned at different resonant
frequencies.
}
The formalism
permits to easily consider filters of any order with an arbitrary
number of nonlinear harmonics. As detailed below, this approach
also {\color{black} sheds light on} the underlying mechanisms that enable
non-reciprocal responses in time-modulated filters. Besides
filters operating at identical input/output frequencies, this
technique can also be applied to analyze devices that exhibit
non-reciprocity between the fundamental frequency and desired
nonlinear harmonics.

{\color{black}
After a review of the equivalent network for coupled resonators
filters in Section~\ref{sec_eqv_network},  we introduce
in Section~\ref{sec_network_time_mod_res}
the coupling matrix formalism for time-modulated
filters. Numerical studies are first presented, including the convergence
behavior of the scattering parameters with the number of harmonics.
To demonstrate the usefulness of the proposed approach,
in Section~\ref{sec_num_results} we present the
design of two non-reciprocal filters of third and fourth orders.
The filters are then experimentally demonstrated
in Section~\ref{sec_pract_realization} using quarter wavelength
resonators implemented in microstrip technology.
}
Coupled microstrip
lines are terminated with varactors on one side to build time modulated
resonators. A compact structure is obtained by integrating the
feeding network of the modulating signal in the same board as
the filter. This is achieved by using a coplanar waveguide
feeding network in the ground plane of the device. Numerical
results obtained with the theory presented in this paper show good
agreement with respect to measurements obtained from the
manufactured prototypes.
\section{Equivalent Network of Coupled Resonators Filters}
\label{sec_eqv_network}
Let us start from the basic ideal equivalent network of a
lossless in-line filter represented by lumped elements and admittance
inverters as shown in Fig.~\ref{fig:eq_circuit_reciprocal}.
{\color{black}
Fig.~\ref{fig:eq_circuit_reciprocal}(a) shows the normalized lowpass
filter prototype with all capacitors normalized to ($1$F) and
the source and load impedances normalized to ($1\Omega$).
For the sake of clarity, but without loss of generality,
we consider a network composed of three resonators (network of order $N=3$).
The ($N+2$) coupling matrix can be used to characterize
this network \cite{cameronlibrogeneric}, leading to
\begin{equation}
\underline{\underline{M}}=
\begin{pmatrix}
0    & M_{{P_1}1} & 0      & 0      & 0 \\
M_{{P_1}1} & M_{11} & M_{12} & 0      & 0 \\
0    & M_{12} & M_{22} & M_{23} & 0  \\
0    & 0      & M_{23} & M_{33} & M_{3{P_2}}  \\
0    & 0      & 0      & M_{3{P_2}} & 0
\end{pmatrix}.
\end{equation}
Here we have used the notation ($P_1$) and ($P_2$) to refer to
the source ($S$) and load ($L$) terminations. This notation will be more
convenient when investigating the non-reciprocal
behavior of the network in the next section. Note that in this
matrix the diagonal elements ($M_{u,u}$ with $u=1,2, \cdots, N$)
represent the frequency invariant susceptances shown in
Fig.~\ref{fig:eq_circuit_reciprocal}(a), while the off-diagonal
elements ($M_{u,u+1}$) represent the values of the admittance inverters
of the network.
Frequency invariant susceptances are used
in Fig.~\ref{fig:eq_circuit_reciprocal} to account for
asynchronously tuned topologies \cite{cameronlibrogeneric}.
\begin{figure}
\centering
\includegraphics[width=9.1cm]{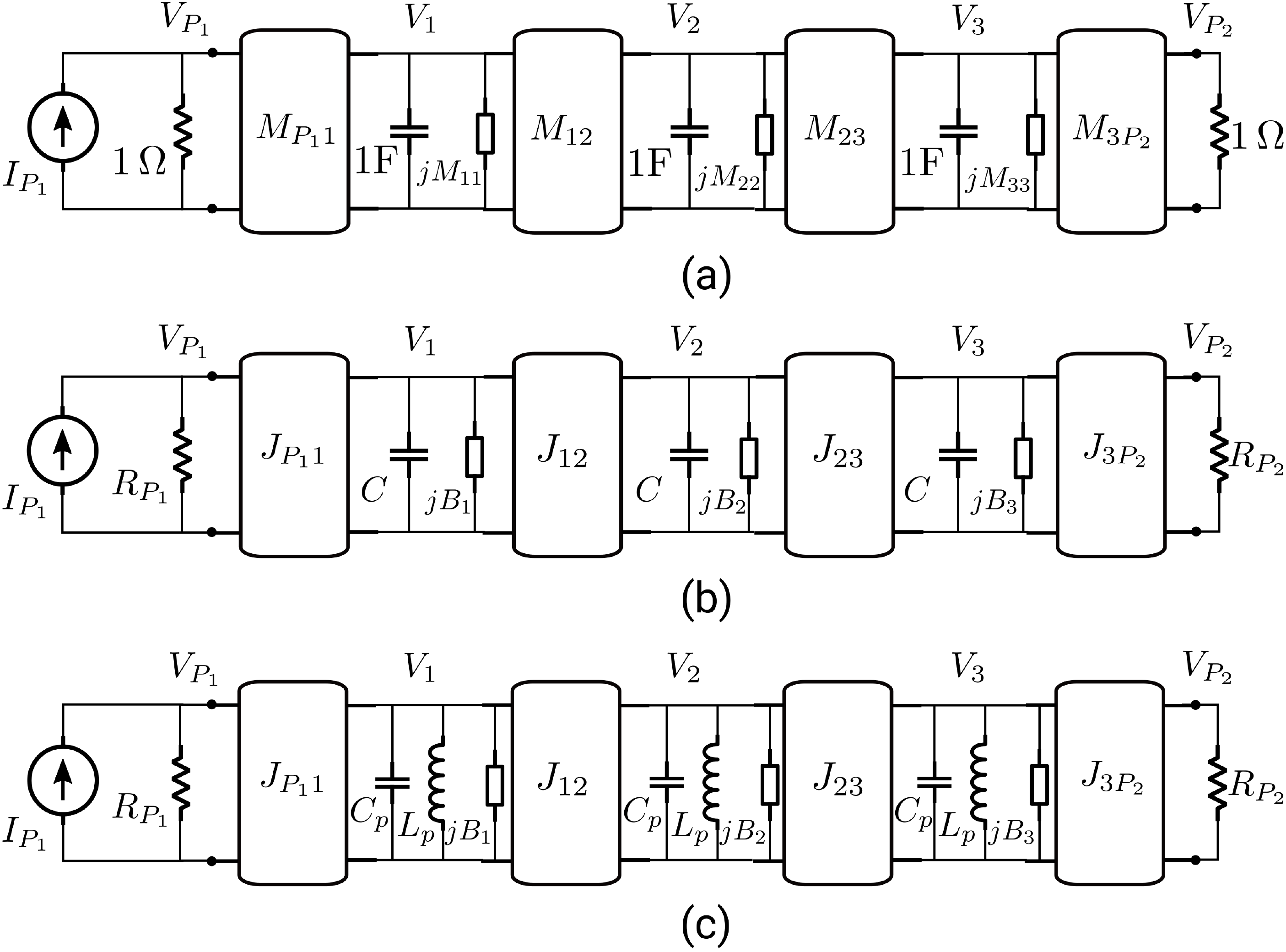}
\caption{Equivalent circuit of an ideal lossless filter based on lumped
	elements and admittance inverters. (a) Normalized lowpass
	prototype with all elements having unitary values.
	(b) Lowpass prototype scaled
	to arbitrary capacitance values ($C$) and port impedances
	($R_{P_1}$, $R_{P_2}$). (c) Bandpass network
	resulting from a standard
	lowpass to bandpass transformation.}
\label{fig:eq_circuit_reciprocal}
\end{figure}

This coupling matrix relates the currents and nodal voltages
in the normalized network
shown in Fig.~\ref{fig:eq_circuit_reciprocal}(a). The
Kirchhoff's current law in this network can be written
in matrix form as
\begin{equation}
 \underline{I} = \Bigl[
	 \underline{\underline{G}}+
	 j \, \omega \, \underline{\underline{C}}+
	 j \, \underline{\underline{M}}
	 \Bigr] \cdot \underline{V},
\end{equation}
where the whole admittance matrix has been expressed as the
sum of three simpler matrices. In this expression
($\underline{\underline{C}}$) is a matrix containing
the capacitors of the network
\begin{equation}
\underline{\underline{C}} =
\begin{pmatrix}
	0   & 0 & 0 & 0 & 0 \\
	0   & C & 0 & 0 & 0 \\
	0   & 0 & C & 0 & 0 \\
	0   & 0 & 0 & C & 0 \\
	0   & 0 & 0 & 0 & 0
\end{pmatrix}
\end{equation}
and ($\underline{\underline{G}}$) is the so called conductance
matrix, which contains the port admittances as
\begin{equation}
\label{eq_cond_matrix}
\underline{\underline{G}} =
\begin{pmatrix}
	G_{P_1} & 0 & 0 & 0 & 0 \\
	0   & 0 & 0 & 0 & 0 \\
	0   & 0 & 0 & 0 & 0 \\
	0   & 0 & 0 & 0 & 0 \\
	0   & 0 & 0 & 0 & G_{P_2}
\end{pmatrix},
\end{equation}
and for the network in Fig.~\ref{fig:eq_circuit_reciprocal}(a):
$C=1$F, $G_{P_1}=1/R_{P_1}=1\, \Omega^{-1}$,
$G_{P_2}=1/R_{P_2}=1\, \Omega^{-1}$.
In this system of equations ($\underline{I}$) represents the
current excitation vector and ($\underline{V}$) contains the unknown
nodal voltages [see Fig.~\ref{fig:eq_circuit_reciprocal}(a)], as
\begin{equation}
\underline{I}=
\begin{pmatrix}
	I_{P_1} \\
	0       \\
	0       \\
	0       \\
	0
\end{pmatrix}, \quad
\underline{V}=
\begin{pmatrix}
	V_{P_1}  \\
	V_1      \\
	V_2      \\
	V_3      \\
	V_{P_2}
\end{pmatrix}.
\end{equation}

From this normalized network, a scaled lowpass circuit as shown
in Fig.~\ref{fig:eq_circuit_reciprocal}(b) can be obtained.
Capacitors and port impedances are scaled to arbitrary
values ($C$), and ($R_{P_1}=1/G_{P_1}$, $R_{P_2}=1/G_{P_2}$), respectively.
{\color{black}
Note that during the production of a particular filter, the transformation
ratios ($C$) are calculated with the information of the practical
technology that will be used during the filter implementation.
}
In any case, the response of the scaled network is the same as the
original network if the values of the admittance inverters ($J_{u,u+1}$)
and frequency invariant susceptances ($B_u$) are conveniently
scaled as
\begin{subequations}
\label{eq:inverters_reciprocal}
\begin{align}
\label{eq:inverters}
J_{{P_1}1}& =M_{{P_1}1} \, \sqrt{G_{P_1} \, C}, & J_{u,u+1}& =M_{u,u+1}\, C, \\
\label{eq:fir}
J_{N{P_2}}& =M_{N{P_2}} \, \sqrt{C \, G_{P_2}}, & B_u&= M_{u,u} \, C,
\end{align}
\end{subequations}
}

If a standard lowpass to bandpass transformation is applied to the network
of Fig.~\ref{fig:eq_circuit_reciprocal}(b), the capacitors are transformed
into resonators, thus obtaining the traditional bandpass network
shown in Fig.~\ref{fig:eq_circuit_reciprocal}(c). In this network all
resonators are equal and take the values
\begin{subequations}
\label{eq_trans_ratios}
\begin{align}
C_p&= \dfrac{C}{\omega_0 \, F_B}, &
L_p& =\dfrac{F_B}{\omega_0 \, C}=\dfrac{1}{\omega_0^2 \, C_p}, \\
F_B&=\dfrac{\omega_{c2}-\omega_{c1}}{\omega_0}, & \omega_0& =2 \, \pi \, f_0,
\end{align}
\end{subequations}
where $f_0$ is the center frequency of the passband
and $\omega_{c1}$ and $\omega_{c2}$ are the lower and upper
angular {\color{black} equi-ripple cut-off frequencies
of the passband}, respectively.

\begin{figure}[!t]
\centering
\includegraphics[width=8.9cm]{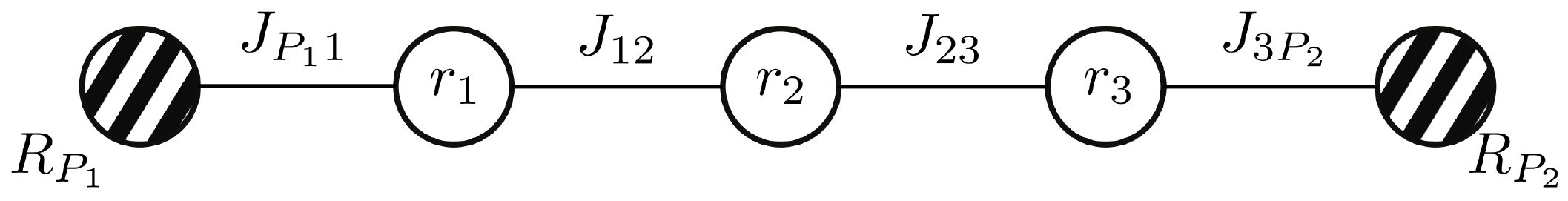}
\caption{Coupling topology of the in-line filter shown
	in Fig.~\ref{fig:eq_circuit_reciprocal}.}
\label{fig:coupling_topology_reciprocal}
\end{figure}

The network shown in Fig.~\ref{fig:eq_circuit_reciprocal}(c) represents
a bandpass filter with the so called in-line coupling topology,
as illustrated in Fig.~\ref{fig:coupling_topology_reciprocal}.
In this figure, white circles represent the resonators of the
structure ($r_u$), while dashed circles represent the terminal ports
with reference impedances ($R_{P_1}=1/G_{P_1}$, $R_{P_2}=1/G_{P_2}$).
Also, solid lines connecting the circles represent the ideal
admittance inverters of the network
($J_{{P_1}1}$, $J_{u,u+1}$, $J_{N{P_2}}$).

If Kirchhoff's current law is applied to the nodes of the bandpass
network shown in Fig.~\ref{fig:eq_circuit_reciprocal}(c), the
following linear system of equations is obtained
\begin{equation}
\label{eq:system_reciprocal}
\begin{pmatrix}
	I_{P_1} \\
	0   \\
	0   \\
	0   \\
	0
\end{pmatrix} =
\begin{pmatrix}
	G_{P_1}     & j J_{{P_1}1}  &    0     &  0       &          0  \\
	j J_{{P_1}1}& Y_p^{(1)} & j J_{12} &  0       &          0  \\
	0       & j J_{12}  & Y_p^{(2)}& j J_{23} &          0  \\
	0       & 0         &j J_{23}  & Y_p^{(3)}& j J_{3{P_2}}    \\
	0       & 0         & 0        &j J_{3{P_2}}  & G_{P_2}
\end{pmatrix} \cdot
\begin{pmatrix}
	V_{P_1} \\
        V_1 \\
        V_2 \\
	V_3 \\
	V_{P_2}
\end{pmatrix}
\end{equation}
where $Y_p^{(u)}$ is the admittance of
the resonators, calculated as
\begin{equation}
\label{eq:ypu_nonmodulated}
Y_p^{(u)} = j \, \omega \, C_p + \dfrac{1}{j \, \omega \, L_p} +
j \, B_u .
\end{equation}
Similarly as before,
it is now convenient to express the matrix of the system
as the sum of three matrices as
\begin{equation}
\label{eq:system_compact}
  \underline{I} = \Bigl[
  \underline{\underline{G}}+
  \underline{\underline{Y_{inv}}}+
  \underline{\underline{Y_p}}
  \Bigr] \cdot \underline{V} .
\end{equation}
The first matrix is again the conductance matrix
defined in~(\ref{eq_cond_matrix}).
The second matrix is symmetric and contains the values of the
admittance inverters of the network
\begin{equation}
\label{eq:yinv}
\underline{\underline{Y_{inv}}} = j \,
\begin{pmatrix}
0     & J_{{P_1}1}&0      & 0    &   0   \\
J_{{P_1}1}& 0     &J_{12} & 0    &   0   \\
0     &J_{12} & 0     &J_{23}&   0   \\
0     &0      &J_{23} & 0    &J_{3{P_2}} \\
0     &0      &0      &J_{3{P_2}}& 0     \\
\end{pmatrix}.
\end{equation}
Finally, the third matrix represents the admittances of the resonators
\begin{equation}
\label{eq:yp}
\underline{\underline{Y_p}} =
\begin{pmatrix}
   0   & 0         & 0         & 0         & 0 \\
   0   & Y_p^{(1)} & 0         & 0         & 0 \\
   0   & 0         & Y_p^{(2)} & 0         & 0 \\
   0   & 0         & 0         & Y_p^{(3)} & 0 \\
   0   & 0         & 0         & 0         & 0
\end{pmatrix}.
\end{equation}
Note that the size of all these matrices is the same as that of the
regular coupling matrix with ports, namely $(N+2) \times (N+2)$.
Also, we want to remark that the admittance inverters are located
in the off diagonal elements of~(\ref{eq:yinv}), and that
the information of the resonators appears in the diagonal
entries of~(\ref{eq:yp}). We stress that all matrices
involved in the formulation are symmetric, therefore assuring
that the considered network is completely reciprocal.

\section{Network with Time Modulated Resonators}
\label{sec_network_time_mod_res}
Applying time-varying signals to modulate the capacitors of the
bandpass network shown in Fig.~\ref{fig:eq_circuit_reciprocal}
makes the system non-linear \cite{pipes54,dikshit73}. In this work,
we will consider that the values of the capacitors are modulated in
time with the following sinusoidal variation
\begin{equation}
\label{eq:mod_capacitor}
   C_p^{(u)}(t) = C_p\, \Bigl[
   1 + \Delta_m \, \cos(\omega_m \, t + \varphi_u) \Bigr],
\end{equation}
where $\omega_m$ is the angular frequency of the modulating
signal, $\varphi_u$ is the initial phase, and $\Delta_m$ is the
modulation index. Even though we will use the same modulation
frequency and modulation index to modulate all capacitors,
their initial phases may be different along the network,
i.e., $\varphi_u=(u-1)\, \Delta_{\varphi}$ with $u=1,2,\cdots, N$.
It will be shown later in this paper that this phase difference
is the key mechanism that enables non-reciprocal responses \cite{wu18}.

\begin{figure}[!t]
\centering
\includegraphics[width=9.0cm]{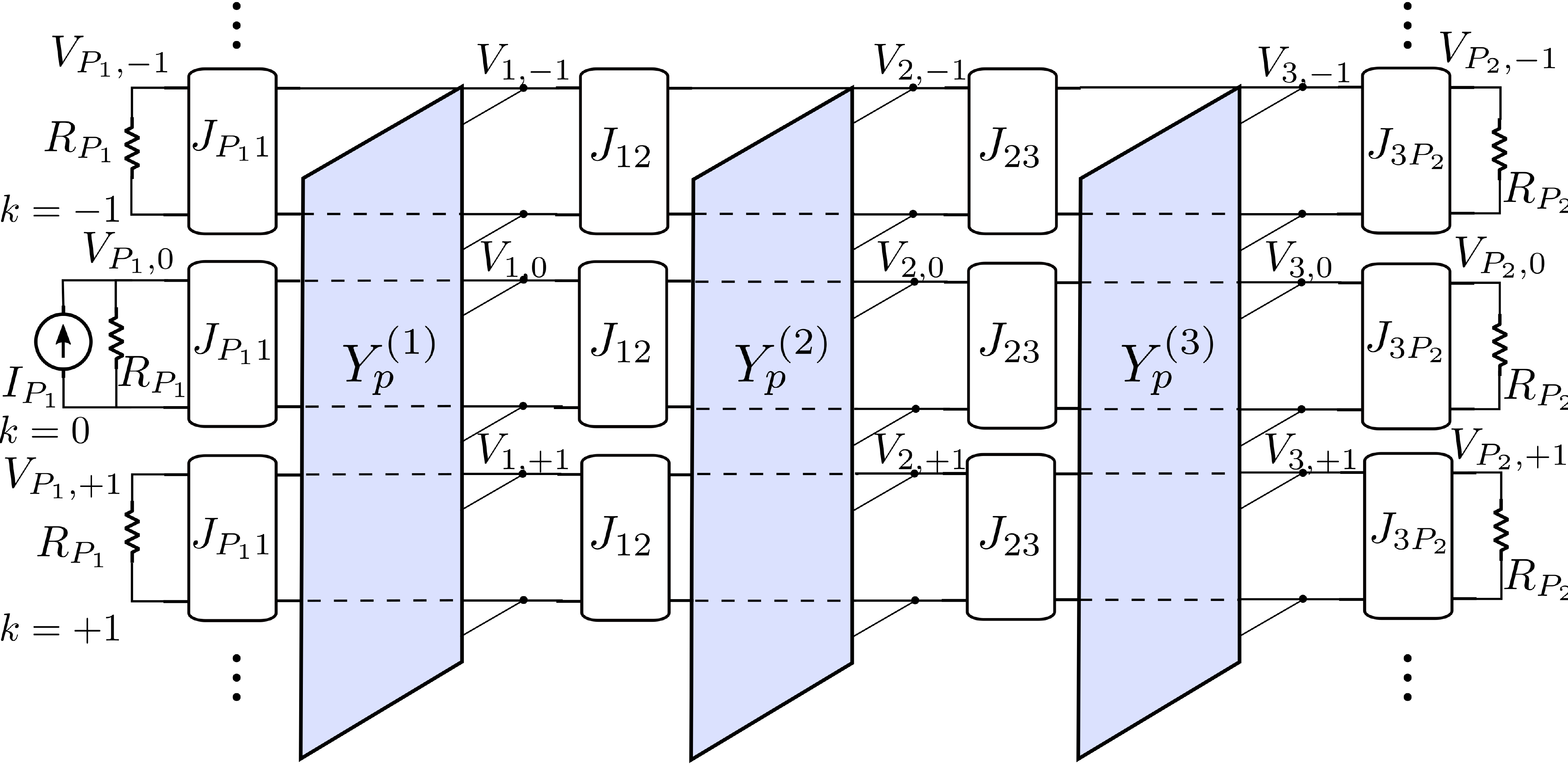}
\caption{Equivalent circuit of the ideal filter shown in
	Fig.~\ref{fig:eq_circuit_reciprocal}(c) when a time domain signal is
	used to modulate the value of the capacitors. Color boxes
	represent the admittance coupling matrix between generated harmonics.}
\label{fig:eq_circuit_nonreciprocal}
\end{figure}

In this scenario, a number of nonlinear
harmonics (or intermodulation products)
$N_{har}$ are generated in each resonator, resulting
into the equivalent network shown in Fig.~\ref{fig:eq_circuit_nonreciprocal}.
These nonlinear harmonics are coupled by the time modulated capacitors.
For simplicity, the figure only shows $N_{har}=3$ harmonics
(i.e., $k=\cdots,-1,0,1,\cdots$ with $k$ denoting the order of a
given nonlinear harmonic).

The application of Kirchhoff's current law on the network shown
in Fig.~\ref{fig:eq_circuit_nonreciprocal} leads to a linear system
with a structure very similar to the previous one given
in~(\ref{eq:system_compact}). However, each entry in the matrix
system becomes now a submatrix of size $N_{har}\times N_{har}$
due to the generated nonlinear harmonics. In this way, the vector
containing the nodal voltages becomes
\begin{subequations}
\label{eq_vol_harmonics}
\begin{align}
\label{eq_vol_harmonics_a}
\underline{V}&=
   \begin{pmatrix}
   \underline{V_{P_1}} \\
   \underline{V_1} \\
   \underline{V_2} \\
   \underline{V_3} \\
   \underline{V_{P_2}}
\end{pmatrix}, &
\underline{V_{P_1}}&=
\begin{pmatrix}
   V_{P_1,-2} \\
   V_{P_1,-1} \\
   V_{P_1,0} \\
   V_{P_1,+1} \\
   V_{P_1,+2}
\end{pmatrix}, \\
\label{eq_vol_harmonics_b}
\underline{V_u}&=
\begin{pmatrix}
   V_{u,-2} \\
   V_{u,-1} \\
   V_{u,0} \\
   V_{u,+1} \\
   V_{u,+2}
\end{pmatrix}, &
\underline{V_{P_2}}&=
\begin{pmatrix}
   V_{P_2,-2} \\
   V_{P_2,-1} \\
   V_{P_2,0} \\
   V_{P_2,+1} \\
   V_{P_2,+2}
\end{pmatrix},
\end{align}
\end{subequations}
where the number of harmonics considered is
five ($N_{har}=5$, $k=\cdots,-2,-1,0,1,2,\cdots$) and the total
number of unknowns in the system of linear equations
becomes $(N+2) \, N_{har}$.
{\color{black}
We recall that in our notation ($u$) is an integer sweeping the
physical resonators ($u=1,2, \cdots N$). Then~($\underline{V_u}$)
of~(\ref{eq_vol_harmonics_b}) are simply
the $2$ to $N+1$ entries
of ($\underline{V}$) shown in~(\ref{eq_vol_harmonics_a}).
}
Then, following the same strategy as before,
the conductance matrix is written as
\begin{equation}
\underline{\underline{G}} =
\begin{pmatrix}
   \underline{\underline{G_{P_1}}} & \underline{\underline{0}} &
   \underline{\underline{0}}   & \underline{\underline{0}} &
   \underline{\underline{0}} \\
   \underline{\underline{0}} & \underline{\underline{0}} &
   \underline{\underline{0}} & \underline{\underline{0}} &
   \underline{\underline{0}} \\
   \underline{\underline{0}} & \underline{\underline{0}} &
   \underline{\underline{0}} & \underline{\underline{0}} &
   \underline{\underline{0}} \\
   \underline{\underline{0}} & \underline{\underline{0}} &
   \underline{\underline{0}} & \underline{\underline{0}} &
   \underline{\underline{0}} \\
   \underline{\underline{0}} & \underline{\underline{0}} &
   \underline{\underline{0}} & \underline{\underline{0}} &
   \underline{\underline{G_{P_2}}}
\end{pmatrix},
\end{equation}
where ($\underline{\underline{0}}$) denotes the zero matrix.
The other sub-matrices are diagonal and represent the loads to the
new generated harmonics as
$\underline{\underline{G_{P_1}}}=G_{P_1} \, \underline{\underline{U}}$
and $\underline{\underline{G_{P_2}}}=G_{P_2}\, \underline{\underline{U}}$,
with $\underline{\underline{U}}$ being the identity matrix.
In addition,
the matrix of the admittance inverters can now be written as
\begin{equation}
\label{eq:yinv_final}
\underline{\underline{Y_{inv}}} = j \,
\begin{pmatrix}
   \underline{\underline{0}} & \underline{\underline{J_{P_{1}1}}} &
   \underline{\underline{0}} & \underline{\underline{0}} &
   \underline{\underline{0}}   \\
   \underline{\underline{J_{P_{1}1}}} & \underline{\underline{0}} &
   \underline{\underline{J_{12}}} & \underline{\underline{0}} &
   \underline{\underline{0}}   \\
   \underline{\underline{0}} & \underline{\underline{J_{12}}} &
   \underline{\underline{0}} & \underline{\underline{J_{23}}} &
   \underline{\underline{0}} \\
   \underline{\underline{0}} & \underline{\underline{0}} &
   \underline{\underline{J_{23}}} & \underline{\underline{0}} &
   \underline{\underline{J_{3P_{2}}}} \\
   \underline{\underline{0}} & \underline{\underline{0}} &
   \underline{\underline{0}} & \underline{\underline{J_{3P_{2}}}} &
   \underline{\underline{0}}
\end{pmatrix},
\end{equation}
where the submatrices
$\underline{\underline{J_{u,u+1}}}=J_{u,u+1}\, \underline{\underline{U}}$
are also diagonal and represent the couplings of same order harmonics
between the different resonators.
{\color{black}
Here we should remark that with the equivalent network employed, which
uses ideal frequency independent inverters, the couplings
of same order harmonics between different resonators are all
identically affected by the original inverters. This is
a narrowband approximation, usually introduced in the theory
of coupling matrices \cite{cameronlibrogeneric}.
In real implementations, harmonics will
be affected by the inverters in a slightly different way, due to
their intrinsic dispersive nature. These dispersive effects maybe important
for wideband responses, and special techniques may be needed
to preserve accuracy \cite{soto10,vannin04}. However, for narrowband
responses (fractional bandwidths typically less than 10\%),
the narrowband approximation usually gives good results
\cite{cameronlibrogeneric}.
}

Finally, the matrix that contains
the resonator admittances becomes
\begin{equation}
\underline{\underline{Y_{p}}} =
\begin{pmatrix}
   \underline{\underline{0}} & \underline{\underline{0}} &
   \underline{\underline{0}} & \underline{\underline{0}} &
   \underline{\underline{0}} \\
   \underline{\underline{0}} & \underline{\underline{Y_p^{(1)}}} &
   \underline{\underline{0}} & \underline{\underline{0}} &
   \underline{\underline{0}} \\
   \underline{\underline{0}} & \underline{\underline{0}} &
   \underline{\underline{Y_p^{(2)}}} & \underline{\underline{0}} &
   \underline{\underline{0}} \\
   \underline{\underline{0}} & \underline{\underline{0}} &
   \underline{\underline{0}} & \underline{\underline{Y_p^{(3)}}} &
   \underline{\underline{0}} \\
   \underline{\underline{0}} & \underline{\underline{0}} &
   \underline{\underline{0}} & \underline{\underline{0}} &
   \underline{\underline{0}}
\end{pmatrix}.
\end{equation}
Each admittance submatrix represents the coupling among the
different nonlinear harmonics generated in a resonator with
a time-modulated capacitor.
{\color{black}
In this paper we have used the theory reported in \cite{darlington63,kurth77}
to model this non-linear behavior. Note that this theory is
based on considering ideal capacitors.
}
Applying the theory reported
in \cite{darlington63,kurth77} permits to express each of
these submatrices as
\begin{equation}
\label{eq:yp_submatrix}
   \underline{\underline{Y_p^{(u)}}}=
   \underline{\underline{Y_{b}}} +
   j \, \underline{\underline{\omega_n}} \,
   \underline{\underline{N_c^{(u)}}} +
   j \, B_u \,  \underline{\underline{U}},
\end{equation}
where $\underline{\underline{\omega_n}}$ is a diagonal matrix
containing the angular frequencies of the nonlinear
harmonics (spectral matrix), namely
\begin{equation}
\underline{\underline{\omega_n}}=
\begin{pmatrix}
  \omega-2 \omega_m &
  0   & 0   & 0    & 0     \\
  0                 &
 \omega - \omega_m  & 0   & 0        & 0     \\
  0                 & 0   & \omega   & 0    & 0     \\
  0  & 0  & 0 &
  \omega+\omega_m & 0 \\
  0  & 0  & 0 & 0 &
  \omega+2 \omega_m
\end{pmatrix}.
\end{equation}
The matrix $\underline{\underline{Y_{b}}}$ includes the presence
of the inductors in the modulated resonators and can be expressed as
\begin{equation}
\underline{\underline{Y_{b}}} =
	\dfrac{1}{j \, L_p} \, \,
	{\underline{\underline{\omega_n}}}^{-1}.
\end{equation}	
Finally, $\underline{\underline{N_c^{(u)}}}$ models how the
nonlinear harmonics are excited due to the modulated capacitors
and it can be written as
\begin{equation}
  \underline{\underline{N_c^{(u)}}} =
  \begin{pmatrix}
	  C_p & D^{(u)} & 0  &  0  & 0 \\
	  E^{(u)} & C_p & D^{(u)} & 0  &  0  \\
	  0  & E^{(u)} & C_p & D^{(u)} & 0   \\
	  0  &   0  & E^{(u)} & C_p & D^{(u)}   \\
	  0  &   0  &   0   & E^{(u)} & C_p
  \end{pmatrix}.
\end{equation}
The new elements of this matrix depend on the modulation index
and on the phases of the modulating signal as
\begin{align}
	D^{(u)} & = \dfrac{\Delta_m \, C_p}{2} \, e^{-j \, \varphi_u}, &
	E^{(u)} & = \dfrac{\Delta_m \, C_p}{2} \, e^{+j \, \varphi_u}.
\end{align}
By doing straightforward operations with these matrices, the
final admittance submatrix in~(\ref{eq:yp_submatrix}) can
be written as shown in~(\ref{eq:yp_submatrix_final}) (top of the page).
\begin{table*}[!t]
\begin{equation}
\label{eq:yp_submatrix_final}
\underline{\underline{Y_p^{(u)}}} =
\begin{pmatrix}
	Y_{r}^{(-2)}+j B_u & j D^{(u)} \, (\omega - 2 \omega_m) & 0  & 0 & 0 \\
	j E^{(u)} \, (\omega - \omega_m) & Y_{r}^{(-1)}+j B_u &
	j D^{(u)} \, (\omega - \omega_m) & 0  & 0 \\
	0 & j E^{(u)} \, \omega & Y_{r}^{(0)}+j B_u &
	j D^{(u)} \, \omega  & 0   \\
	0  & 0 & j E^{(u)} \, (\omega + \omega_m)  & Y_{r}^{(+1)}+j B_u &
	j D^{(u)} \, (\omega + \omega_m)    \\
	0  &  0  & 0 & j E^{(u)} \, (\omega+2 \omega_m)  &
	Y_{r}^{(+2)}+ j B_u
\end{pmatrix}
\end{equation}
\hrule
\end{table*}
In this last expression, we have employed the following auxiliary admittance
\begin{equation}
\label{eq_yrk}
	Y_r^{(k)} = j \, C_p \, \Bigl( \omega + k \, \omega_m \Bigr) +
	\dfrac{1}{j \, L_p \, \Bigl( \omega + k \, \omega_m \Bigr)}.
\end{equation}

The form of the matrix shown in~(\ref{eq:yp_submatrix_final})
admits an interesting interpretation of the non-linear phenomenon
in terms of coupled network resonators. Following the coupling
matrix formalism, the elements in the diagonal represent new
resonators due to the generated nonlinear harmonics
(we will call them harmonic resonators). Therefore, each physically
modulated resonator gives rise to $N_{har}$ new harmonic
resonators yielding to a network of order $N_{har} \, N$.
These resonators have different resonant frequencies, transforming
the original structure into an asynchronously tuned coupled
resonators network.

The resonant frequencies of the new harmonic resonators can
be obtained by equating the diagonal elements of the matrix
shown in~(\ref{eq:yp_submatrix_final}) to zero. However,
following the coupling matrix formalism, it would be convenient
to formulate all resonators to be equal, with additional
{\color{black} frequency invariant susceptances}
to account for differences in the resonant frequencies.
This can be accomplished by first writing~(\ref{eq_yrk}) as
\begin{equation}
	Y_r^{(k)} = j \, \omega \, C_p +
	j \, C_p \, k \, \omega_m +
	\dfrac{1}{j\, \omega \, L_p \Bigl(1 + k \, \omega_m/\omega \Bigr)},
\end{equation}
and then applying the following Taylor expansion
\begin{equation}
	\dfrac{1}{1+x} \approx 1 -x + \cdots, \qquad x<1
\end{equation}
to the third term to obtain
\begin{equation}
\label{eq:Admittance_approx}
	Y_r^{(k)} \approx j \, \omega \, C_p +
	\dfrac{1}{j \, \omega \, L_p} +
	j \, \Biggl(
	C_p \, k \, \omega_m +
	\dfrac{k \, \omega_m}{\omega^2 \, L_p}
	\Biggr).
\end{equation}
{\color{black}
Note that this Taylor expansion can be used in this context since, in
general, we will assume: $\omega_m<<\omega$. This assumption is
again related to the narrowband approximation assumed throughout
the paper, and to the fact that to achieve good power conversion
between non-linear harmonics, the modulation frequency should
lay within the passband of the filter \cite{wu18}.
}

The comparison of this expression with~(\ref{eq:ypu_nonmodulated})
shows that the harmonic resonators
can be made all equal to the static resonators in the unmodulated
network. The differences in resonant frequencies can be modeled with
additional
{\color{black} frequency invariant susceptances}, defined as
\begin{equation}
	\hat{B}_k=C_p \, k \, \omega_m +
	\dfrac{k \, \omega_m}{\omega_0^2 \, L_p} \, ,
\end{equation}
where, in order to make the
{\color{black} frequency invariant susceptances}
truly independent on frequency,
the center frequency of the passband $\omega_0$ has been used in the
last definition. The approximation will remain valid for
narrowband filters. These
{\color{black} frequency invariant susceptances}
can also be formulated in terms of
the initial lowpass capacitors as
\begin{equation}
\label{eq:fir_harmonic_resonators}
\hat{B}_k=\dfrac{2 \, k \, \omega_m \, C}{\omega_0 \, F_B}.
\end{equation}

It can be observed that the
{\color{black} frequency invariant susceptances associated to}
harmonic resonators
depend on the order of the nonlinear harmonic itself $k$, on
the modulation frequency $\omega_m$ and on the passband bandwidth.
This expression is also very useful, since it will directly
translate into the diagonal elements of the coupling matrix for
the non-reciprocal filter by setting the lowpass capacitor
to unity, i.e., $C=1$.

It is illustrative to compare the structure of the matrices
shown in~(\ref{eq:system_reciprocal}) and
in~(\ref{eq:yp_submatrix_final}). Specifically, the off diagonal
elements of the matrix~(\ref{eq:yp_submatrix_final}) indicate that
the new harmonic resonators are coupled following an in-line coupling
topology among them. However, it can be observed that the matrix
is not symmetric. This indicates that these harmonic resonators
are coupled through non-reciprocal admittance inverters. Following
this idea, we define a non-reciprocal admittance inverter to represent
the coupling between two different harmonics $k-1$ and $k$,
belonging to a specific physical resonator $u$, as
\begin{equation}
\label{eq:inverter_non_reciprocal}
\left\{
\begin{array}{ll}
  J_u^{(k,k-1)} =
  D^{(u)} \, \bigl[ \omega+k \, \omega_m\bigr], &
	  \mbox{Low to up.} \\
  J_u^{(k-1,k)} =
	E^{(u)} \, \bigl[ \omega+(k-1) \, \omega_m\bigr], &
	\mbox{Up to low.} \rule[-0mm]{0mm}{4mm}
\end{array}
\right.
\end{equation}
so a coupling from a lower order harmonic to an upper order harmonic
will use the top formula of~(\ref{eq:inverter_non_reciprocal}),
while a coupling from an upper order harmonic to a lower order harmonic
will involve the bottom formula. An explicit expression for this
non-reciprocal inverter can be obtained in the lowpass domain as
\begin{equation}
\label{eq:inverter_non_reciprocal_lowpass}
\left\{
\begin{array}{l}
  J_u^{(k,k-1)} =
  \dfrac{\Delta_m}{2} \dfrac{C}{\omega_0 \, F_B} \, e^{-j \varphi_u} \,
  \bigl[ \omega_0+k \, \omega_m\bigr] \\
  J_u^{(k-1,k)} =
  \dfrac{\Delta_m}{2} \dfrac{C}{\omega_0 \, F_B} \, e^{+j \varphi_u} \,
        \bigl[ \omega_0+(k-1) \, \omega_m\bigr]
        \rule[-0mm]{0mm}{6mm}
\end{array}
\right.
\end{equation}
where the center angular frequency of the passband has been used to
define frequency invariant inverters.

{\color{black}
Here we remark that these admittance inverters are different
from those shown in~(\ref{eq:yinv_final}). Admittance inverters
in~(\ref{eq:yinv_final}) come from the unmodulated network, and
they couple same order harmonics between different physical
resonators. On the contrary, these new admittance inverters
play an important role in the non-linear process occurring within
each time modulated resonator. As a consequence, the new admittance
inverters in~(\ref{eq:inverter_non_reciprocal_lowpass}) model
the couplings between the different harmonics, generated, due
to the non-linear process,
within the same physical resonator.
}

These last expressions indicate that the coupling between adjacent
harmonic resonators belonging to a specific physical modulated capacitor
can be controlled with the modulation frequency $\omega_m$,
modulation index $\Delta_m$, and initial phase of the modulation
signal $\varphi_u$. Moreover, the degree of non-reciprocity of the
coupling depends on both, the initial phase of the modulating signal
and the modulation frequency. These expressions represent the
values along the off-diagonal elements of the coupling matrix for
the final non-reciprocal filter, once the value of the lowpass
capacitor is set to unity ($C=1$).

\begin{figure}[!t]
\centering
\includegraphics[width=9cm]{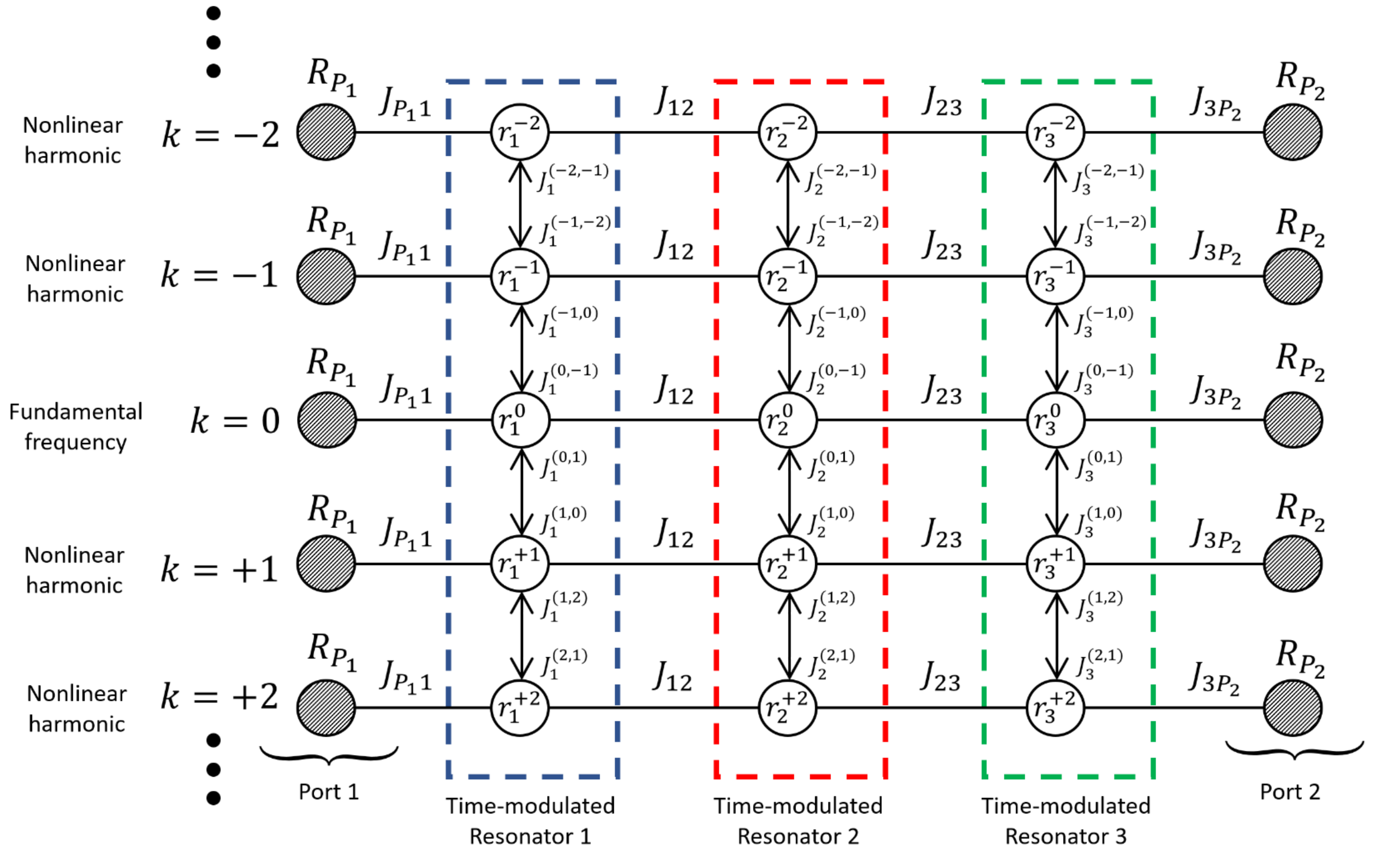}
\caption{Coupling topology of the in-line filter shown in
	Fig.~\ref{fig:eq_circuit_reciprocal}(c) when the capacitors
	of the resonators are modulated with a time varying signal.}
\label{fig:coupling_topology_nonreciprocal}
\end{figure}

The analysis presented above permits an insightful interpretation
of non-reciprocal filters in terms of an asynchronously tuned coupled
resonators network. As already indicated, the order of the
equivalent network is $N \, N_{har}$. Its coupling topology is
further shown in Fig.~\ref{fig:coupling_topology_nonreciprocal}.
In this figure, harmonic resonators are identified with white circles
as $r_u^{(k)}$. These harmonic resonators are defined with the
same inductors ($L_p$), capacitors ($C_p$) and
{\color{black} frequency invariant susceptances}
($B_u$)
as the original static resonators. However, the new
{\color{black} frequency invariant susceptances}
($\hat{B}_k$) given in~(\ref{eq:fir_harmonic_resonators})
must be added to correctly represent their resonant frequencies.
Furthermore, solid lines represent regular inverters modeling the
couplings of same order harmonics between different physical resonators,
as defined in~(\ref{eq:inverters_reciprocal}). Finally, lines
terminated in arrows represent non-reciprocal inverters modeling
the couplings between different order harmonic resonators belonging
to the same physical resonator, as defined
in~(\ref{eq:inverter_non_reciprocal})
or~(\ref{eq:inverter_non_reciprocal_lowpass}). It is also interesting
to note that this coupled resonator network can easily be characterized
with the traditional coupling matrix
formalism \cite{cameronlibrogeneric}, using the results obtained in
this Section. In this case the size of the coupling matrix
is ($N+2) N_{har} \times (N+2) N_{har}$.

\begin{figure}[t]
\centering
\subfloat[]{\includegraphics[width=0.24\textwidth]{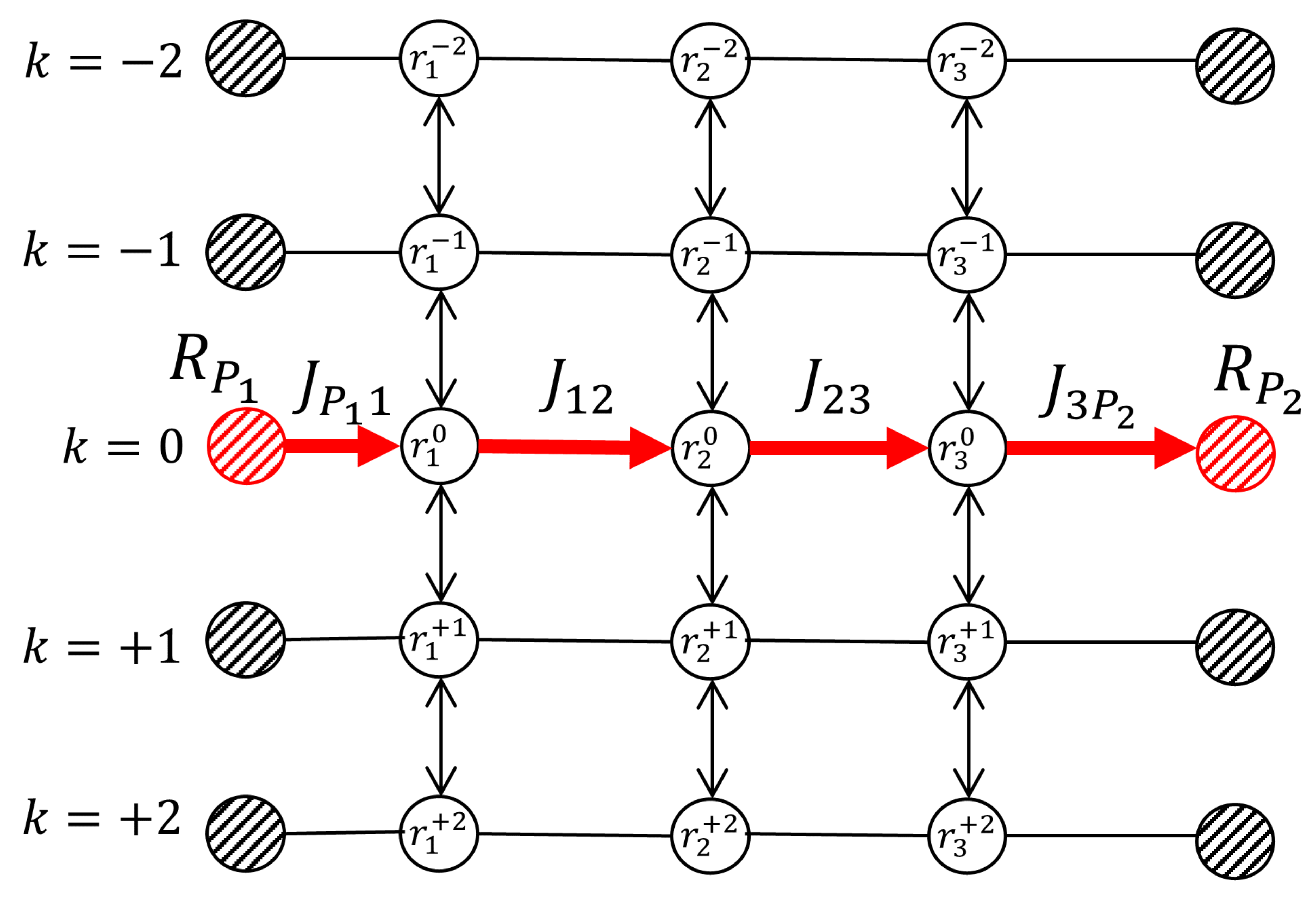}%
\label{fig:Forward_1}
}
\hfill
\subfloat[]{\includegraphics[width=0.23\textwidth]{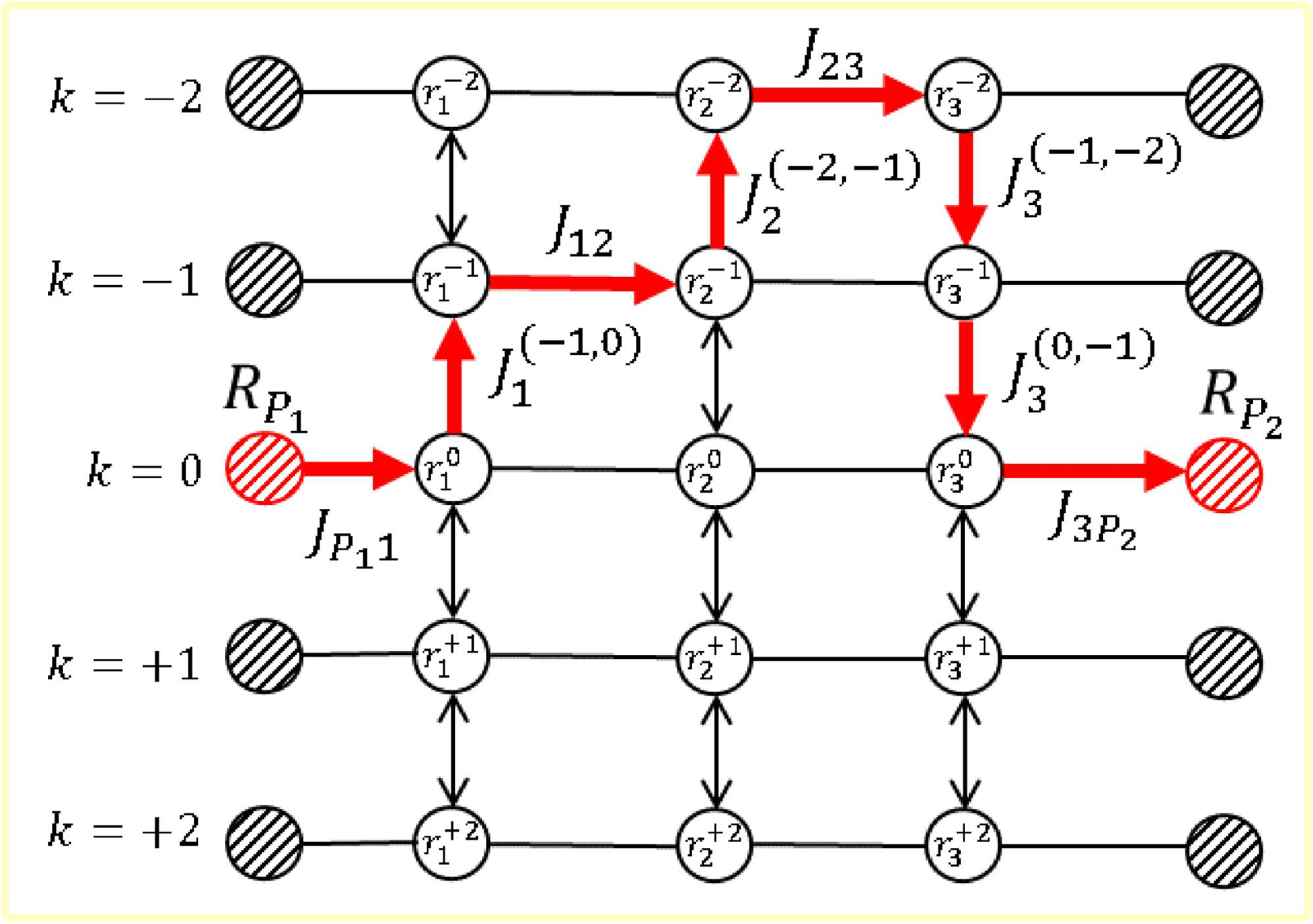}%
\label{fig:Forward_2}
}
\hfill
\subfloat[]{\includegraphics[width=0.23\textwidth]{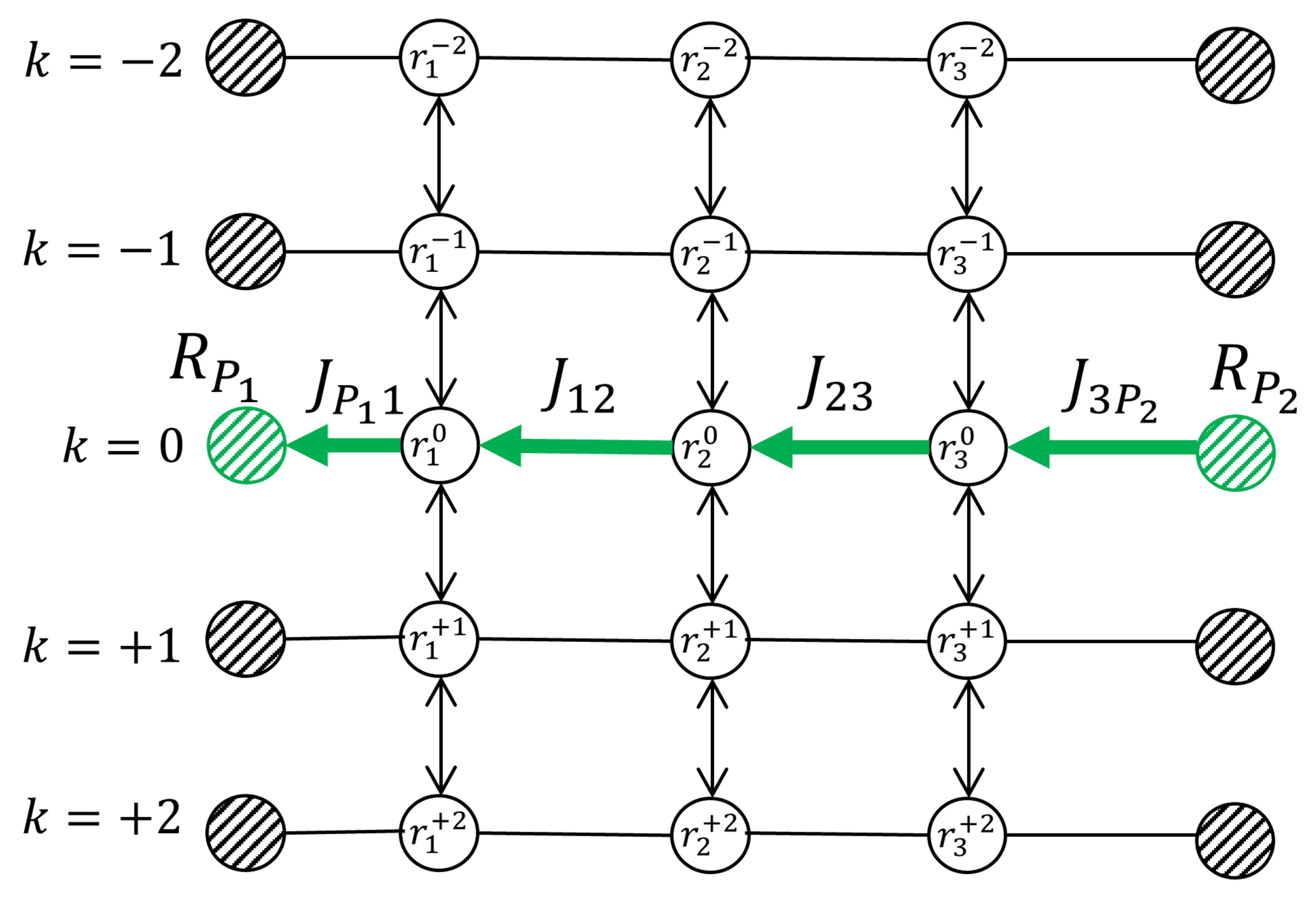}%
\label{fig:Backward_1}
}
\hfill
\subfloat[]{\includegraphics[width=0.23\textwidth]{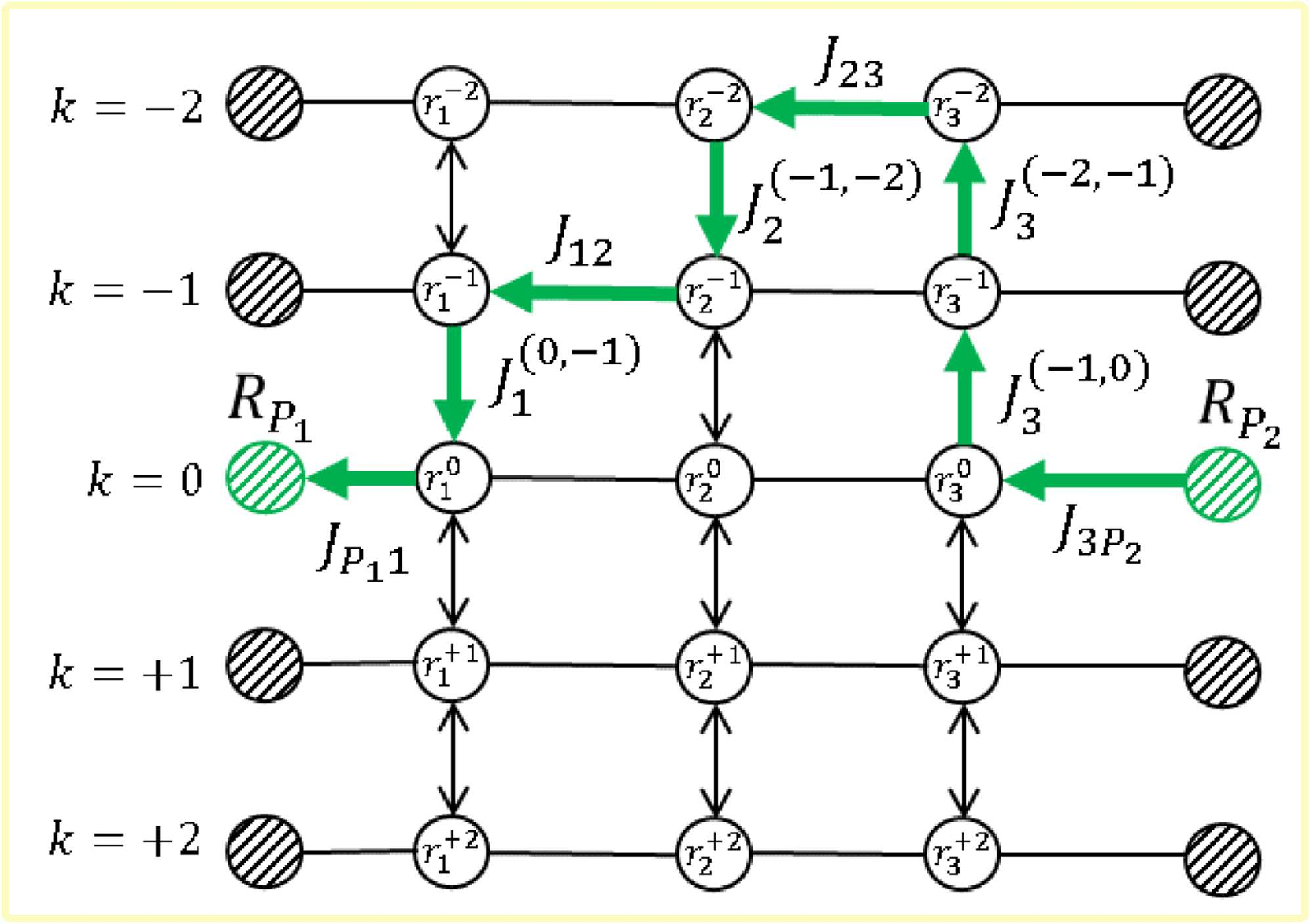}%
\label{fig:Backward_2}
}
\caption{Different paths that can be followed by electromagnetic
	waves to travel from port $1$ to port $2$ (top row) and
	from port $2$ to port $1$ (bottom row) in the coupling topology
	described in Fig.~\ref{fig:coupling_topology_nonreciprocal}.}
\label{fig:Topology_and_path}
\end{figure}

{\color{black}
It is interesting to note that according to the admittance
inverters expressed in~(\ref{eq:inverter_non_reciprocal_lowpass}),
the coupling
increases with the order of the harmonics. This implies that
higher order harmonics will undergo very high couplings, which
is a somewhat counter-intuitive scenario.  The situation,
however, can be explained with the coupling topology shown
in Fig.~\ref{fig:coupling_topology_nonreciprocal}.
This topology explicitly states that couplings to higher
harmonics can only occur from contiguous
harmonics. Therefore, the power cannot be coupled from the fundamental
frequency to harmonics of very high orders, with a very strong
coupling.
}

The topology shown in Fig.~\ref{fig:coupling_topology_nonreciprocal}
explicitly shows that the non-reciprocal response in time-modulated
filters originates due to the non-reciprocal coupling
[see~(\ref{eq:inverter_non_reciprocal_lowpass})] between
adjacent nonlinear harmonics that appear in time-modulated resonators.
Following this scheme, the underlying non-reciprocal mechanism can
be intuitively understood as follows. Electromagnetic waves
propagating from port $1$ can reach port $2$ and keep the
same oscillation frequency by (i) going through the admittance
inverters that link the different resonators at the fundamental frequency,
as in regular in-line filters
(see Fig.~\ref{fig:coupling_topology_reciprocal}
and Fig.~\ref{fig:Forward_1}); and (ii) going through an
ideally infinite number of routes (assuming an infinite number
of nonlinear harmonics) that appear in the topology due to
the presence of harmonic resonators. One specific example of
these routes, illustrated in Fig.~\ref{fig:Forward_2}, involves
the harmonic admittance inverters $J^{(-1,0)}_{1}$,
$J^{(-2,-1)}_{2}$, $J^{(-1,-2)}_{3}$, and $J^{(0,-1)}_{3}$ that
impart a total phase of $+\varphi_1+\varphi_2-2 \varphi_3$ to the
waves propagating therein. The output at port $2$ is then
conformed by the interference of the waves coming from all
possible routes. Let us now consider the dual case, i.e.,
waves coming from port $2$ and propagating towards port $1$.
As in our previous analysis, propagating waves can follow the
path of common in-line filters (see Fig.~\ref{fig:Backward_1})
plus potentially any of the ideally infinite routes enabled
by harmonic resonators. The former leads to reciprocal contributions
whereas any of the paths that encompasses nonlinear harmonics
introduces non-reciprocity due to the non-reciprocal response of the
impedance inverters. For instance, Fig.~\ref{fig:Backward_2} shows
the route previously analyzed but considering now the opposite
propagation direction of the waves. This specific path involves the
same
harmonic impedance inverters as before, but traversed in the opposite
direction, thus providing a total
phase of $-\varphi_1-\varphi_2+2 \varphi_3$ to the waves
(negative with respect
to the previous scenario). For instance, assuming
$\Delta_{\varphi}=45^\circ$, the total phase difference
between forward and backward paths in this example is of $90^\circ$.
It is thus evident that an adequate
control of the phase imparted by each time-modulated resonator is
key to control the response of this type of filters. At port $1$,
waves coming from all routes interfere to construct the output signal.
Strong non-reciprocity at the same frequency arises due to the
different wave interference that appears in ports $1$ and $2$.

The design of time-modulated non-reciprocal filters can be carried
out following the guidelines shown in \cite{wu18}. In such design,
the goal is to optimize the modulation frequency and index as well
as the initial phase of the modulation signal applied to each
resonator to (i) independently manipulate the interference of all
waves that merge at ports $1$ and $2$ to boost non-reciprocity;
(ii) maximize the energy coupled to nonlinear harmonics; and (iii)
ensure that most energy is transferred back to the operation
frequency at the device ports to minimize loss. It is important
to remark that it is required to modulate at least two physical
resonators to enable non-reciprocal responses \cite{wu18}. If one modulates
just a single resonator, the incoming energy will simply be
distributed among various nonlinear harmonics that will then
propagate through the network. Finally, note that we have
{\color{black} focused on}
non-reciprocal responses at the same frequency. It is indeed
possible to design devices based on time-modulated resonators
that exhibit non-reciprocal responses between the fundamental
frequency and any desired nonlinear harmonic. These devices
will be governed by the topology shown in
Fig.~\ref{fig:coupling_topology_nonreciprocal} and will follow
the theory developed here.
\section{Numerical Results}
\label{sec_num_results}
Using the coupling matrix formalism derived above, a software tool
for the analysis of non-reciprocal in-line filters has been developed.
In this Section, we will investigate the convergence of the numerical
algorithm as a function of the number of harmonics $N_{har}$ included
in the calculations.

The first example is a filter of order three whose unmodulated
response has equal ripple return losses of $RL=13$~dB. The filter
coupling matrix yields
\begin{equation}
\label{eq:M_3order}
  \underline{\underline{M_3}} =
  \begin{pmatrix}
	  0       & 0.8894 & 0      &     0  &     0 \\
	  0.8894  & 0      & 0.8294 &     0  &     0 \\
	  0       & 0.8294 & 0      & 0.8294 &     0 \\
	  0       & 0      & 0.8294 & 0      &0.8894 \\
	  0       & 0      & 0      & 0.8894 &     0
  \end{pmatrix}.
\end{equation}
{\color{black}
This coupling matrix gives the response of the normalized
lowpass prototype.
}
The bandpass response is adjusted to have a bandwidth of 47~MHz,
with a center frequency of $f_0=975$~MHz ($F_B=4.8 \%$). By using
the procedure shown in \cite{wu18}, the modulation parameters
were optimized, leading to the following
values: $f_m=22.8$~MHz, $\Delta_m=0.050$, and $\Delta_\varphi=35^\circ$.

{\color{black}
Here we should remark that the design of this filter is
not yet completely determined by synthesis techniques.
Rather, the coupling matrix shown in~(\ref{eq:M_3order})
gives the initial response of the unmodulated filter.
Once this response is established, the parameters of the modulation
signals are optimized to obtain the desired non-reciprocal
response, using the procedure reported in \cite{wu18}.

In general, the design of this filter fully from synthesis techniques
is very complex, and will involve (i) the calculation of
suitable reflection and transmission polynomials
to properly represent the desired
(non-reciprocal) transfer functions, (ii) the extraction from
these polynomials of a suitable coupling matrix and
(iii) the transformation of the obtained coupling matrix
into a form that represents the coupling topology shown
in Fig.~\ref{fig:coupling_topology_nonreciprocal}.
}

\begin{figure}[!t]
\centering
\subfloat[]{\includegraphics[width=0.23\textwidth]{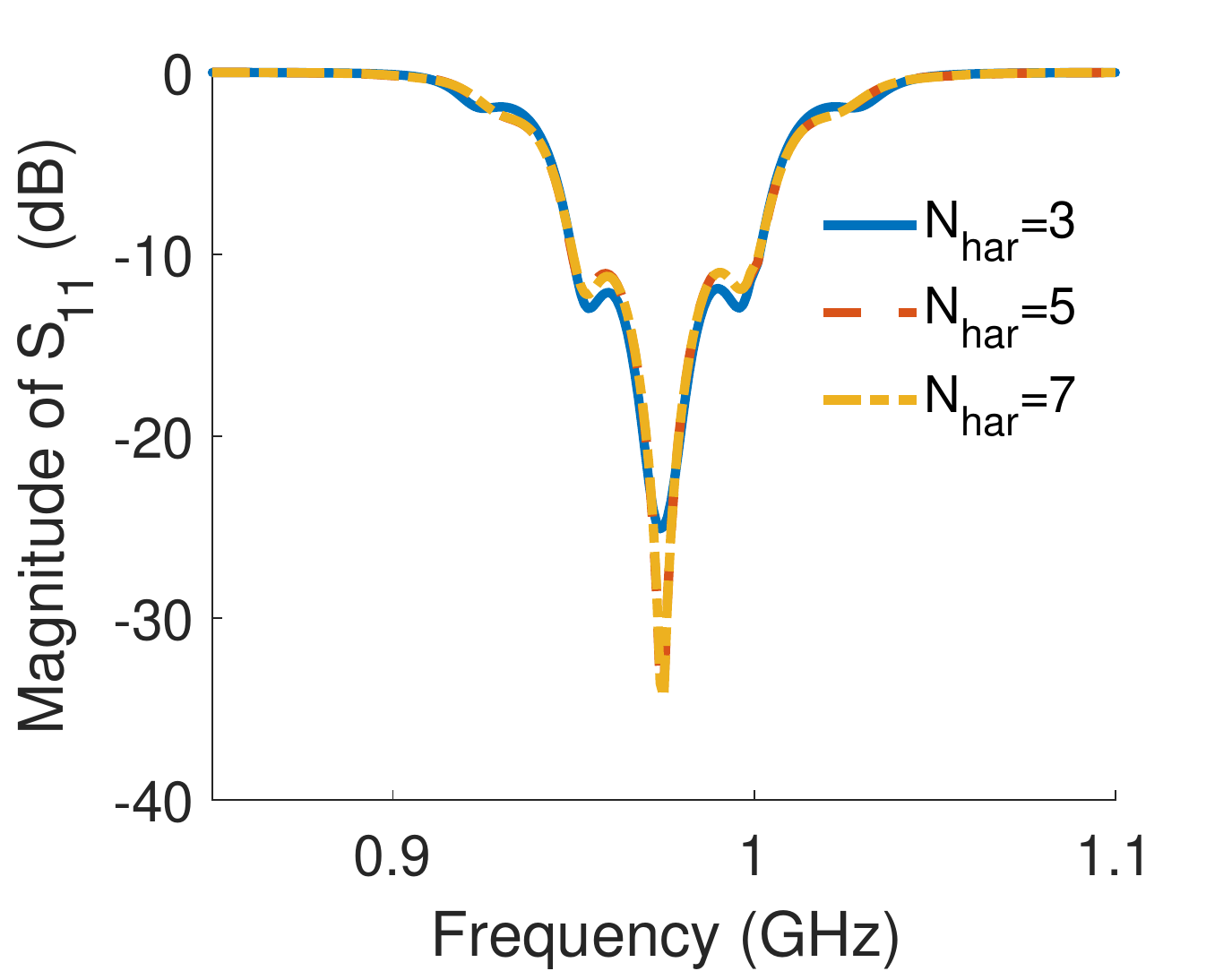}%
\label{fig:S11_Simul_Order_3}
}
\hfill
\centering
\subfloat[]{\includegraphics[width=0.23\textwidth]{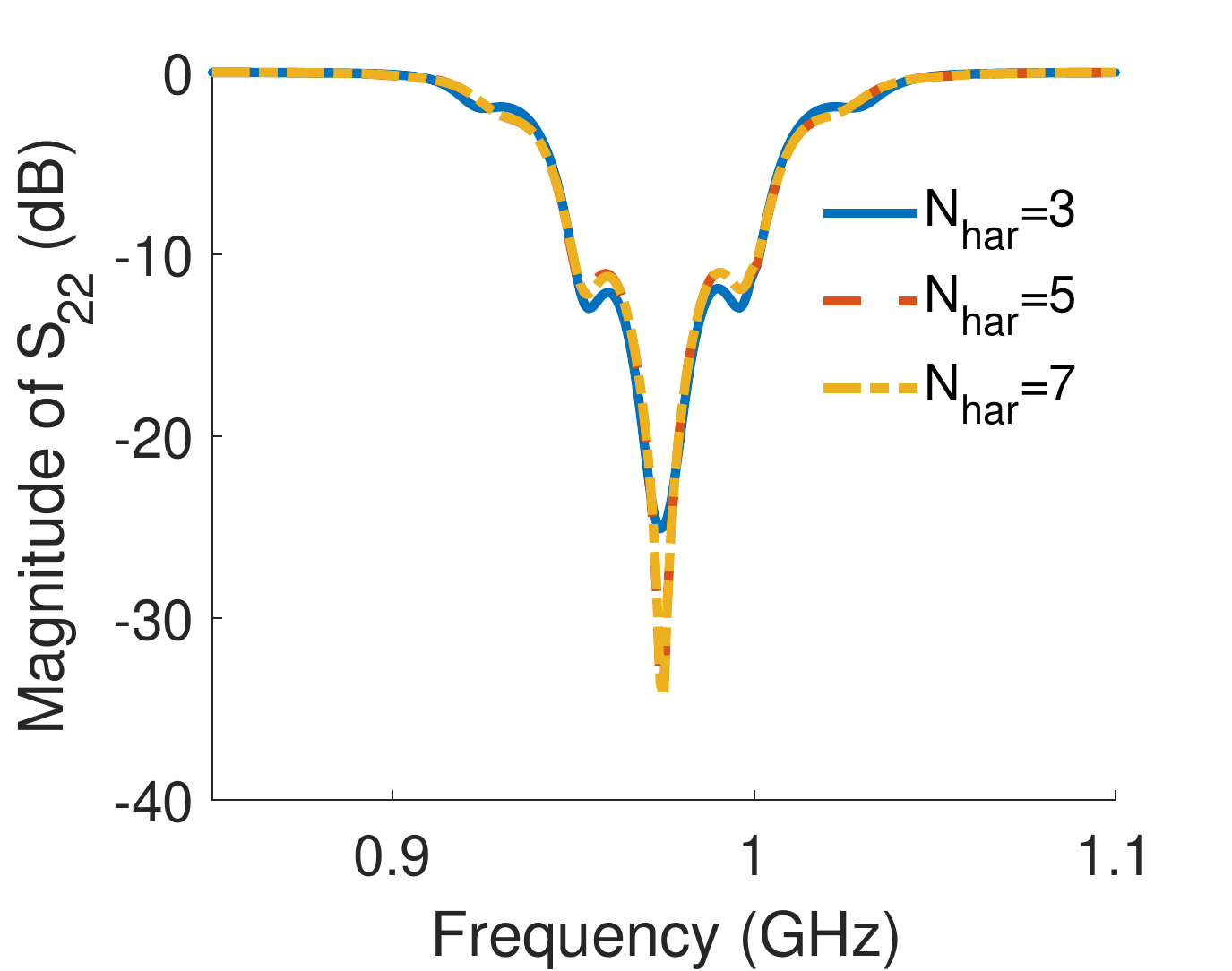}%
\label{fig:S22_Simul_Order_3}
}
\hfill
\subfloat[]{\includegraphics[width=0.23\textwidth]{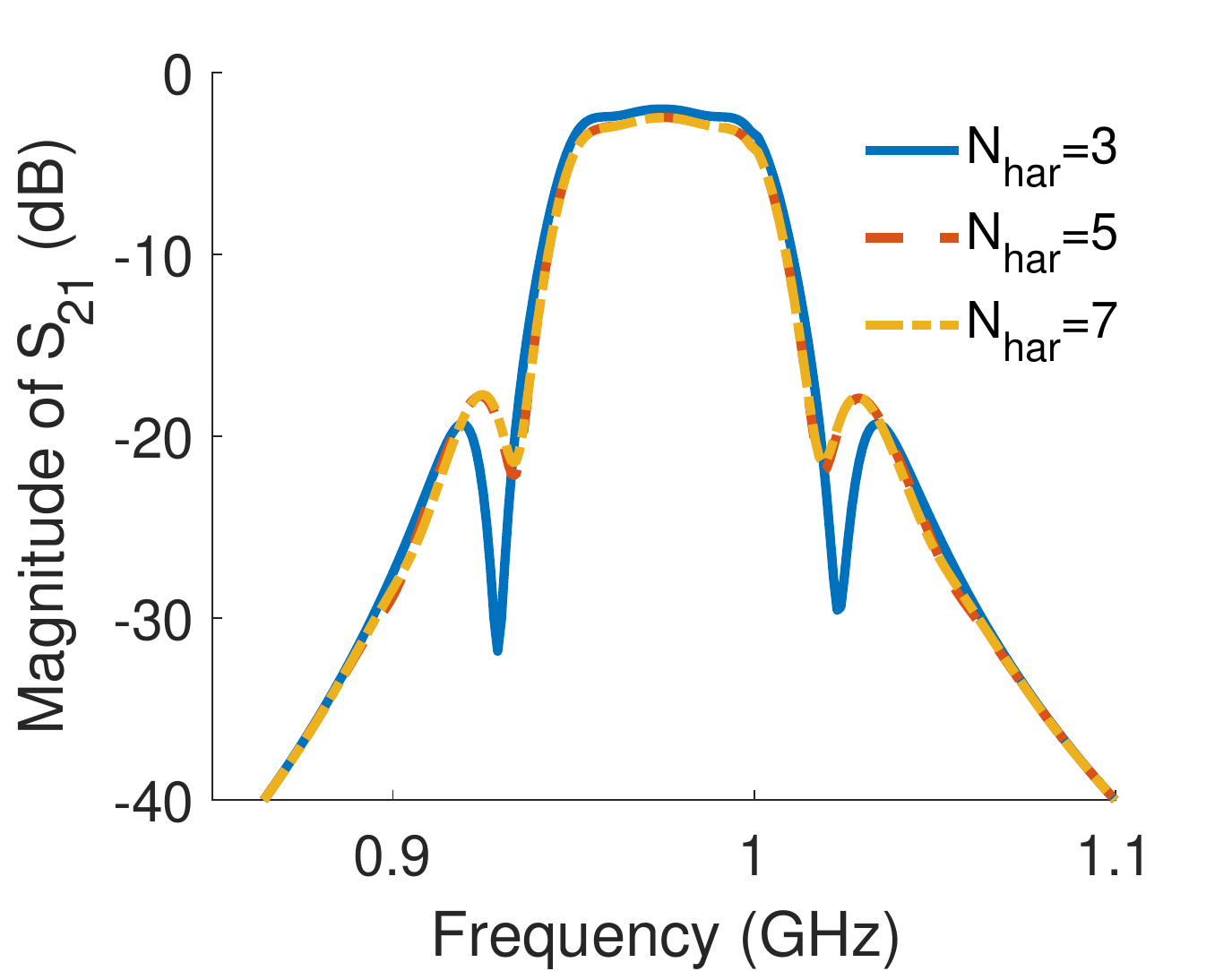}%
\label{fig:S21_Simul_Order_3}
}
\hfill
\subfloat[]{\includegraphics[width=0.23\textwidth]{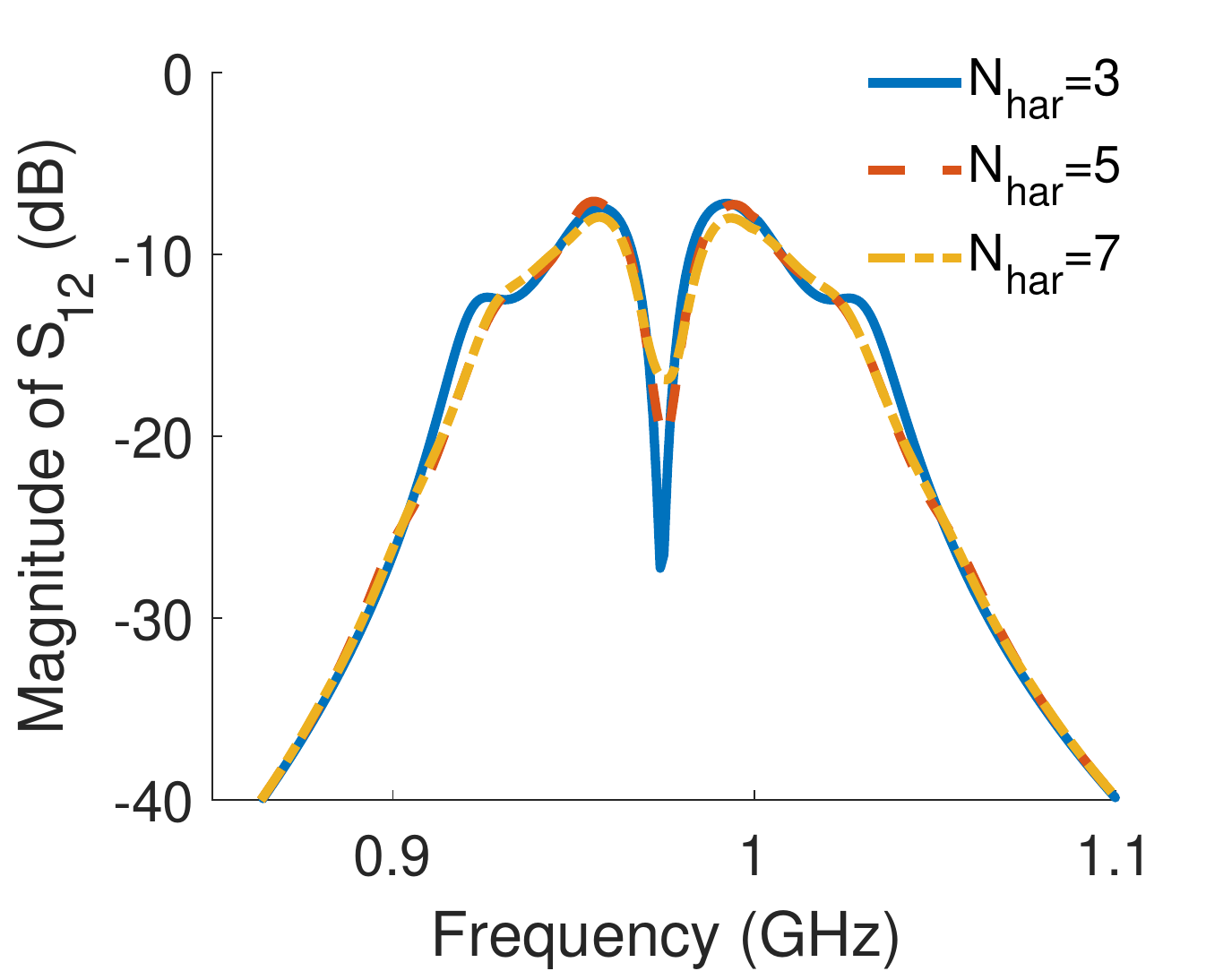}%
\label{fig:S12_Simul_Order_3}
}
\caption{Scattering parameters of the third order non-reciprocal filter
	designed in Section~\ref{sec_num_results}.
	Results are computed with the coupling
	matrix approach introduced in this work using an increasing
	number of harmonics in the numerical method.}
\label{fig:Sparam_Simul_Order_3}
\end{figure}
Fig.~\ref{fig:Sparam_Simul_Order_3} shows the scattering parameters
at the fundamental frequency, obtained for this filter with
increasing number of harmonics $N_{har}=3, \, 5, \, 7$.
{\color{black}
Here numerical results were obtained from the responses
of the coupling matrices for the time modulated network.
The coupling matrix is easily calculated starting with
the coupling matrix given in~(\ref{eq:M_3order}) for the unmodulated
network, and with the selected parameters for the modulation
signal ($f_m$, $\Delta_m$ and $\Delta_{\varphi}$).
Then, using the coupling topology shown in
Fig.~\ref{fig:coupling_topology_nonreciprocal}, the coupling
matrix entries for the time modulated network are computed
with~(\ref{eq:fir_harmonic_resonators})
and~(\ref{eq:inverter_non_reciprocal_lowpass}) (with $C=1$).
}
It can
be observed that the results are in general very stable, showing
only small differences as the number of harmonics is increased.
Note that the algorithm converges using just $N_{har}=5$ harmonics and
increasing further the number of harmonics leads to negligible
changes in the simulated response. Results show that the filter
has a passband which is quite flat in the forward direction with
a bandwidth of $48$~MHz measured at the return loss level of $11$~dB.
It should be stressed that, even though the network is non-reciprocal
it is symmetric and thus return losses from both ports are identical.
Insertion losses within the passband in the forward direction
are $2.5$~dB. Since the network is lossless, these losses are
in fact due to power that is converted to nonlinear harmonics
and is not converted back to the fundamental frequency. A very
strong non-reciprocity is obtained at the center of the passband,
being the insertion loss of about $17$~dB. Overall, the insertion
losses in the backward direction are greater that $8$~dB within
the whole useful bandwidth.
{\color{black}
We observed in this case that fairly good isolation can be obtained
at the center of the passband. However, the isolation deteriorates
at the edges of the useful bandwidth. As explained in the previous
section, the non-reciprocity is obtained by provoking energy
conversion from the fundamental frequency to the generated
non-linear harmonics. Although these conversion effects are
non-reciprocal in magnitude and phase, the main mechanism
that allows to obtain high non-reciprocity is the difference in phase
between the forward and backward paths. Therefore, high isolation
is obtained by adjusting the phases among the resonators to
produce phase cancellation effects in the backward direction.
With a small number of resonators (three in this example),
these cancellation effects
can be made efficient in a narrow bandwidth. Moreover, as it will
be discussed in our next example, there is a trade-off between
the isolation level and the bandwidth where this isolation is
achieved. In general, larger isolation values can be obtained
but only over a narrower bandwidth.
}

If we define the directivity between the forward and backward directions as
\begin{equation}
	D=\dfrac{|S_{21}|^2}{|S_{12}|^2},
\end{equation}
then a directivity of $D_0=14.5$~dB is obtained at center frequency.
Moreover, the directivity within the useful passband is always
better than $D=5.5$~dB.

To demonstrate the convergence of the algorithm when the order of the
network is increased, we have also designed a fourth order non-reciprocal
filter. For this second example the return losses of the unmodulated
filter are $RL=18.5$~dB, leading to the following coupling matrix
\begin{equation}
\label{eq:M_4order}
  \underline{\underline{M_4}} =
  \begin{pmatrix}
       0       & 0.997  & 0      &     0  &     0  &        0  \\
       0.997   & 0      & 0.873  &     0  &     0  &        0  \\
       0       & 0.873  & 0      & 0.68   &     0  &        0   \\
       0       & 0      & 0.68   & 0      & 0.873  &        0  \\
       0       & 0      & 0      & 0.873  &     0  & 0.997     \\
       0       & 0      & 0      & 0      & 0.997  &        0
  \end{pmatrix}.
\end{equation}
This time the bandpass response is adjusted to have a bandwidth
of 58~MHz at a center frequency $f_0=890$~MHz, given a fractional
bandwidth of $F_B=6.5\%$. After optimization, the parameters
of the modulated capacitors are $f_m=19$~MHz, $\Delta_m=0.076$,
and $\Delta_\varphi=48^\circ$.

\begin{figure}[t]
\centering
\subfloat[]{\includegraphics[width=0.23\textwidth]{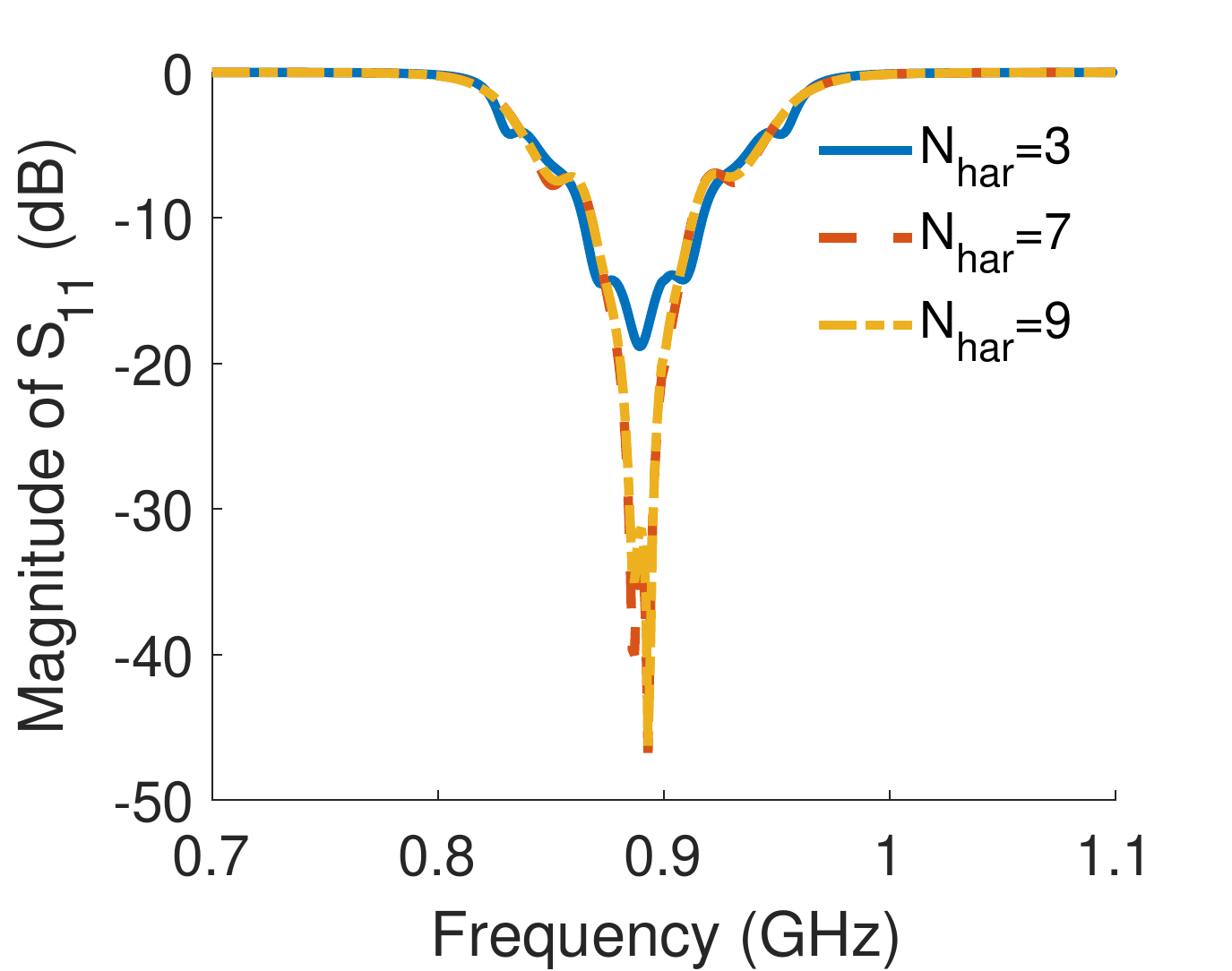}%
\label{fig:S11_Simul_Order_4}
}
\hfill
\subfloat[]{\includegraphics[width=0.23\textwidth]{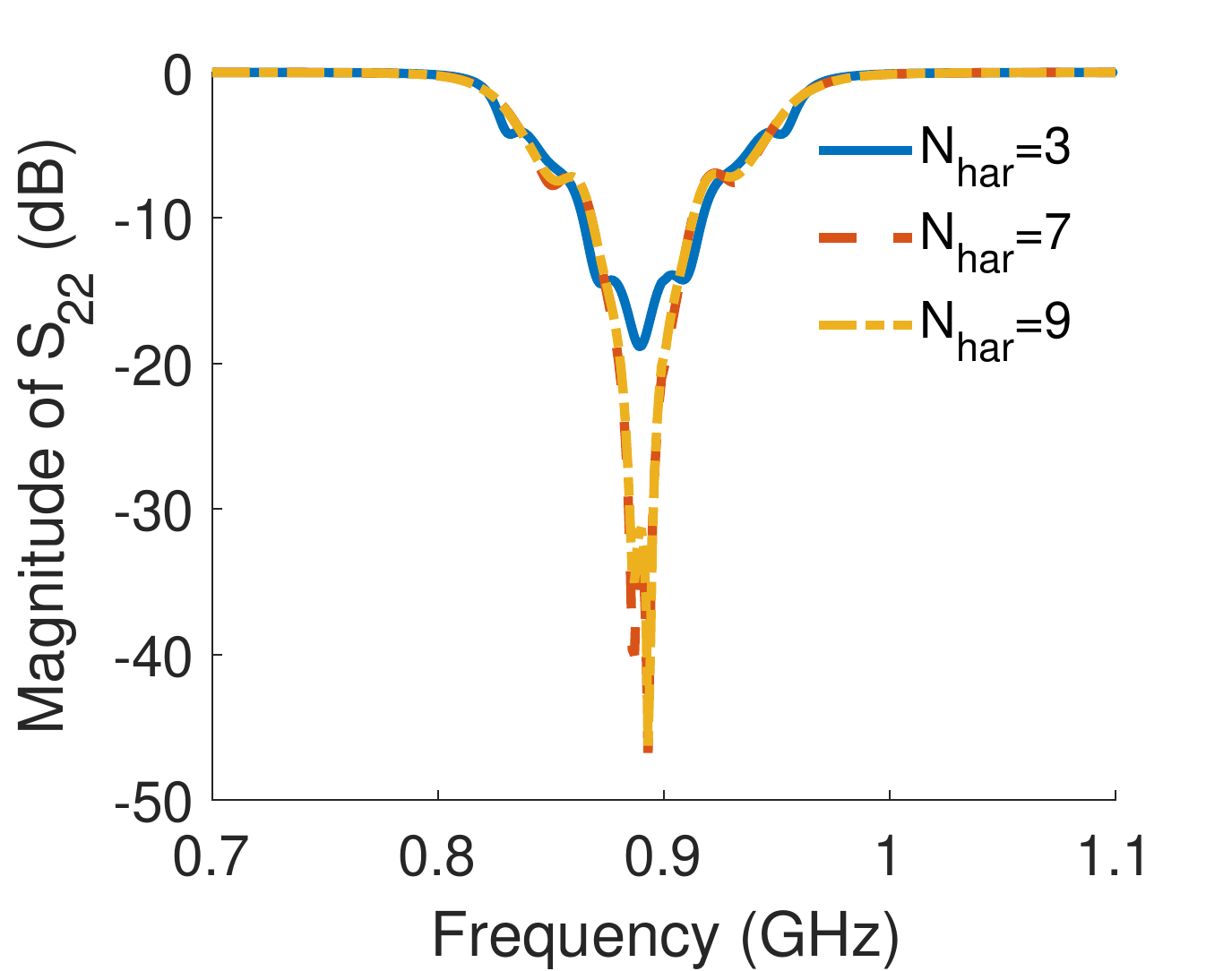}%
\label{fig:S22_Simul_Order_4}
}
\hfill
\subfloat[]{\includegraphics[width=0.23\textwidth]{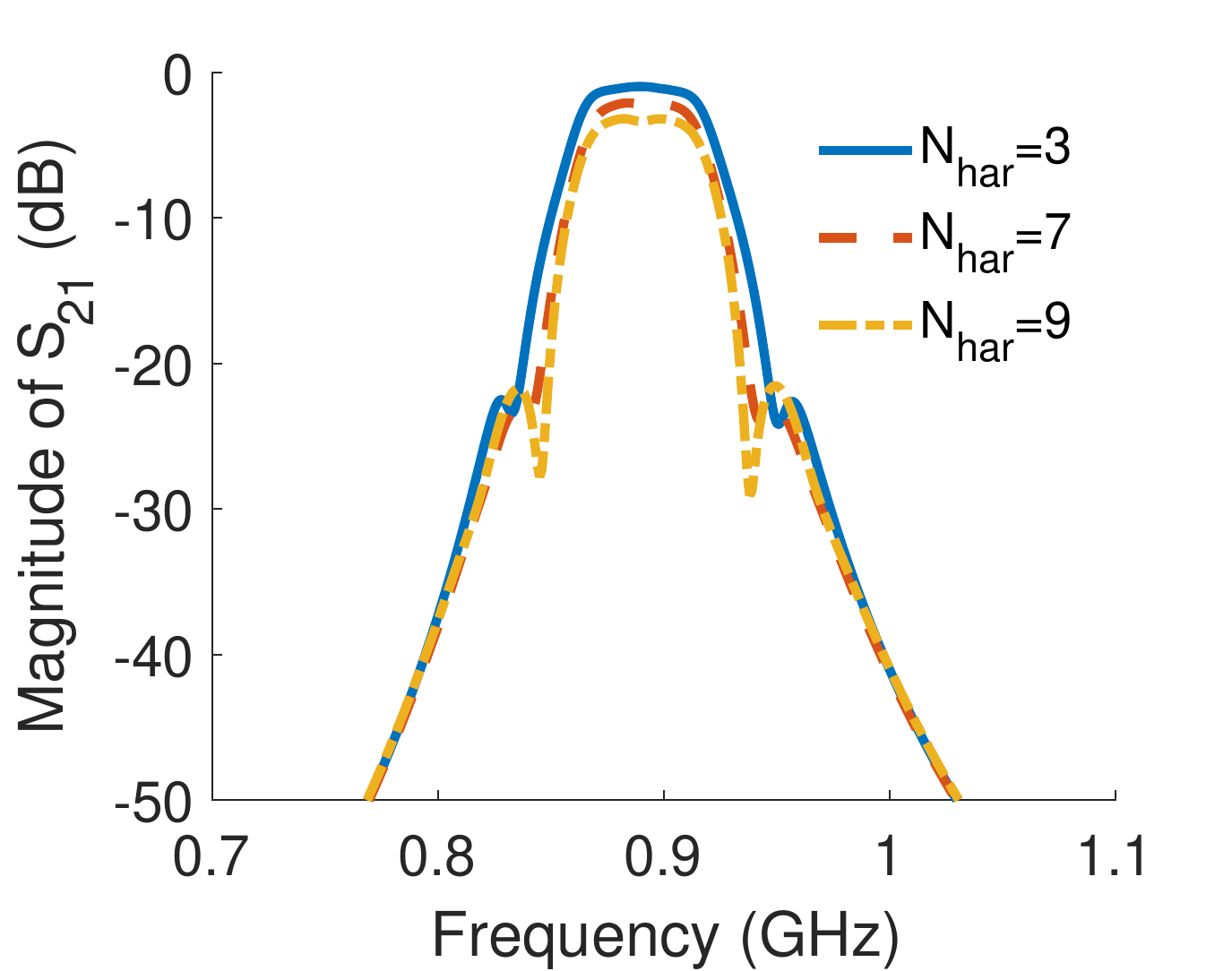}%
\label{fig:S21_Simul_Order_4}
}
\hfill
\subfloat[]{\includegraphics[width=0.23\textwidth]{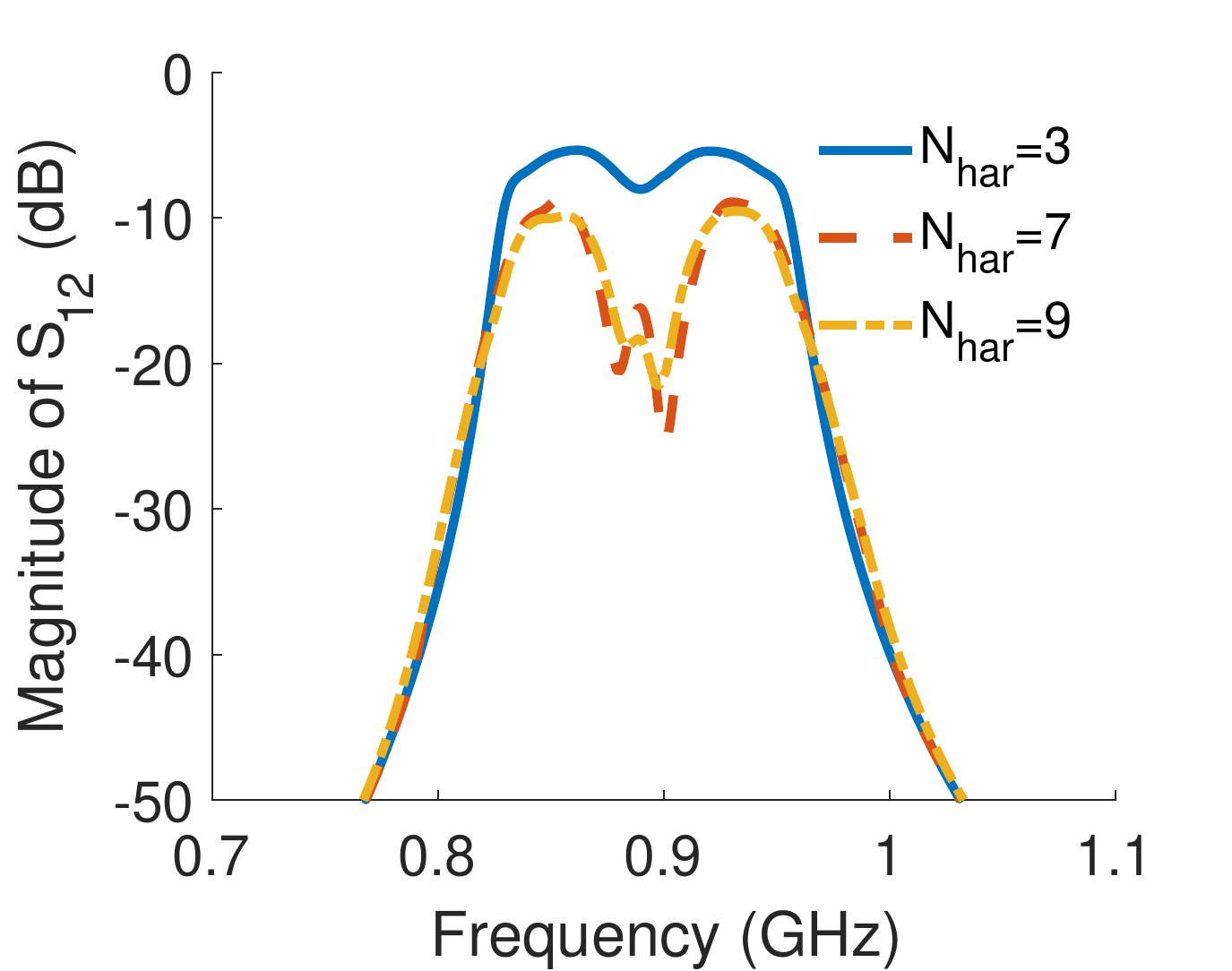}%
\label{fig:S12_Simul_Order_4}
}
\caption{Scattering parameters of the forth order non-reciprocal filter
	designed in Section~\ref{sec_num_results}.
	Results are computed with the
	coupling matrix approach introduced in this work using an
	increasing number of harmonics in the numerical method.}
\label{fig:Sparam_Simul_Order_4}
\end{figure}

Fig.~\ref{fig:Sparam_Simul_Order_4} shows the simulated scattering
parameters with increasing number of nonlinear
harmonics $N_{har}=3, \, 7, \, 9$. It is evident that the
response is inaccurate if only three harmonics are included
in the calculations. After increasing further the number
of harmonics, the differences among the different simulations
reduce considerably, especially for the reflection characteristic
and the forward transmission coefficient. We have verified
that including additional harmonics in the simulations leads
to negligible variations in the simulated response, which
indicates that good convergence is obtained with
nine harmonics. As expected, this study shows that more
harmonics needs to be used in the numerical simulations
when the order of the network increases.

{\color{black}
Moreover, it has been previously shown \cite{qin14},
that in this type of modulated
resonators only the two first higher order
harmonics are important in the non-linear process.
Consequently, the minimum number of harmonics that
need to be considered in the numerical simulations
should grow, with the number of resonators in the
network, according to the rule: $N_{har}=2 \, (N-1)+1$.
Note that the convergence results presented for the
third and fourth order filters, shown in
Fig.~\ref{fig:Sparam_Simul_Order_3} and Fig.~\ref{fig:Sparam_Simul_Order_4},
are in agreement with this rule.
}

The filter shows an almost flat response for the transmission
coefficient in the forward direction, having a bandwidth of $40$~MHz
measured at a return loss of $RL=12$~dB. The insertion losses
in the forward direction are smaller than $IL=3.3$~dB within the useful
passband. Again, these losses correspond to power converted
from the fundamental frequency into nonlinear harmonics that is
not converted back into the fundamental frequency. The response of
the filter shows a strong non-reciprocal behavior in the
backward direction. Around the center frequency, the directivity
is better than $D_0=13.7$~dB in a bandwidth of $26$~MHz.
In the whole useful passband, the directivity is shown to be better
than $D=9$~dB.

{\color{black}
At this point it is interesting to observe that the optimum
modulation frequency ($f_m=19$~MHz) is slightly smaller than
the bandwidth of the filter. This condition assures that
the two first intermodulation products can be strongly excited,
while the generation of higher order intermodulation products
are minimized.
}
{\color{black}
Also, we emphasize that the response of the filter was
optimized to achieve a good trade-off between the isolation
level, and the bandwidth where it is achieved. Other optimization
criteria are possible, for instance by increasing further
the isolation level, at the expense of reducing the
bandwidth where this isolation is achieved. For instance we
have verified that by decreasing the frequency of the modulation
signal to $f_m=18$~MHz, the directivity increases
to $D_0=33.1$~dB, although in a narrow bandwidth of
only $8.6$~MHz.
In any case, this example shows that the proposed system
offers high flexibility in the characteristics that can
be achieved, that could be adapted to many different
scenarios.
}

\begin{figure}[!t]
\centering
\subfloat[]{\includegraphics[width=4.3cm]{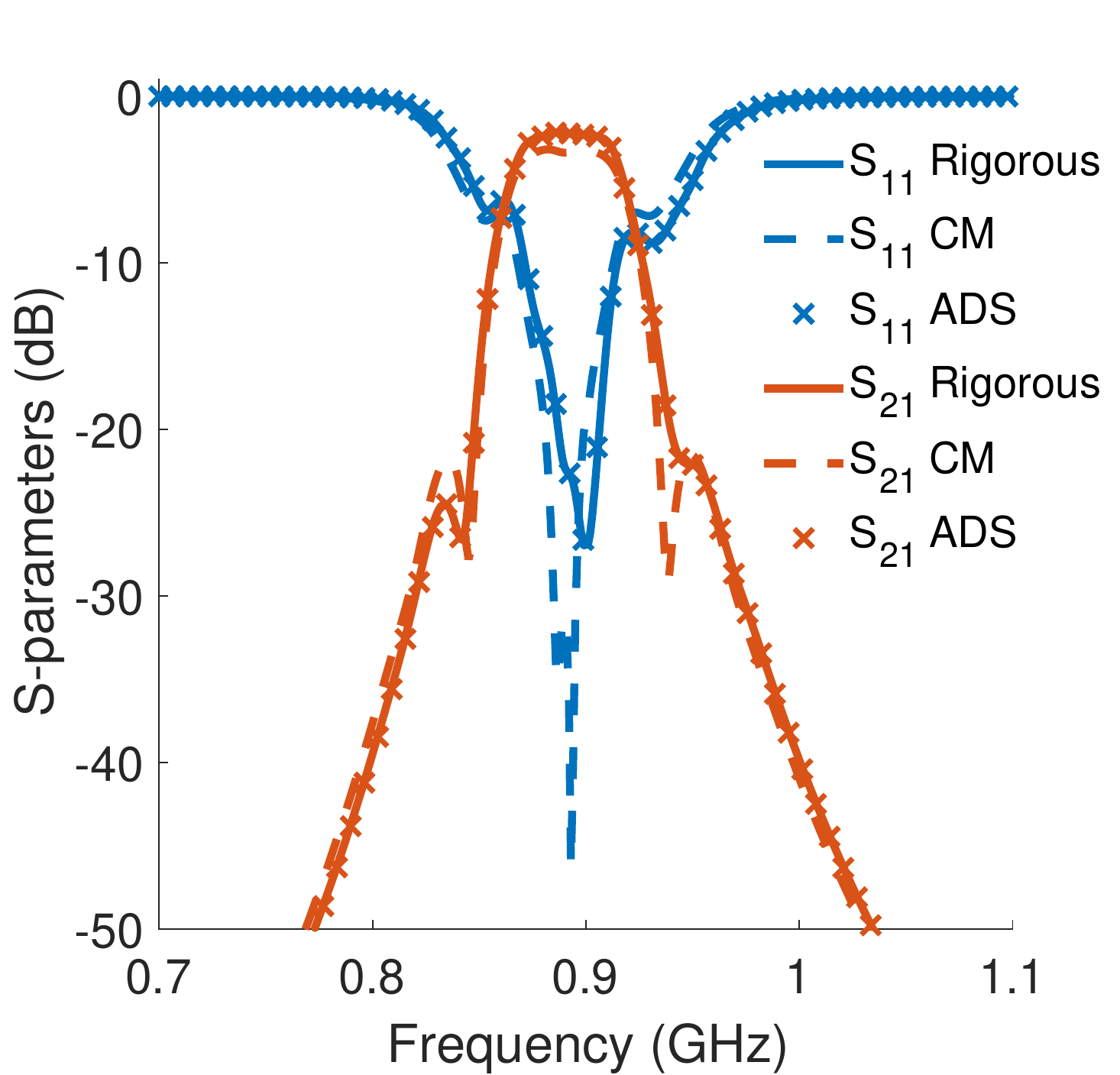}%
\label{fig:Simul_Order_4_Port1}
}
\hfill
\subfloat[]{\includegraphics[width=4.3cm]{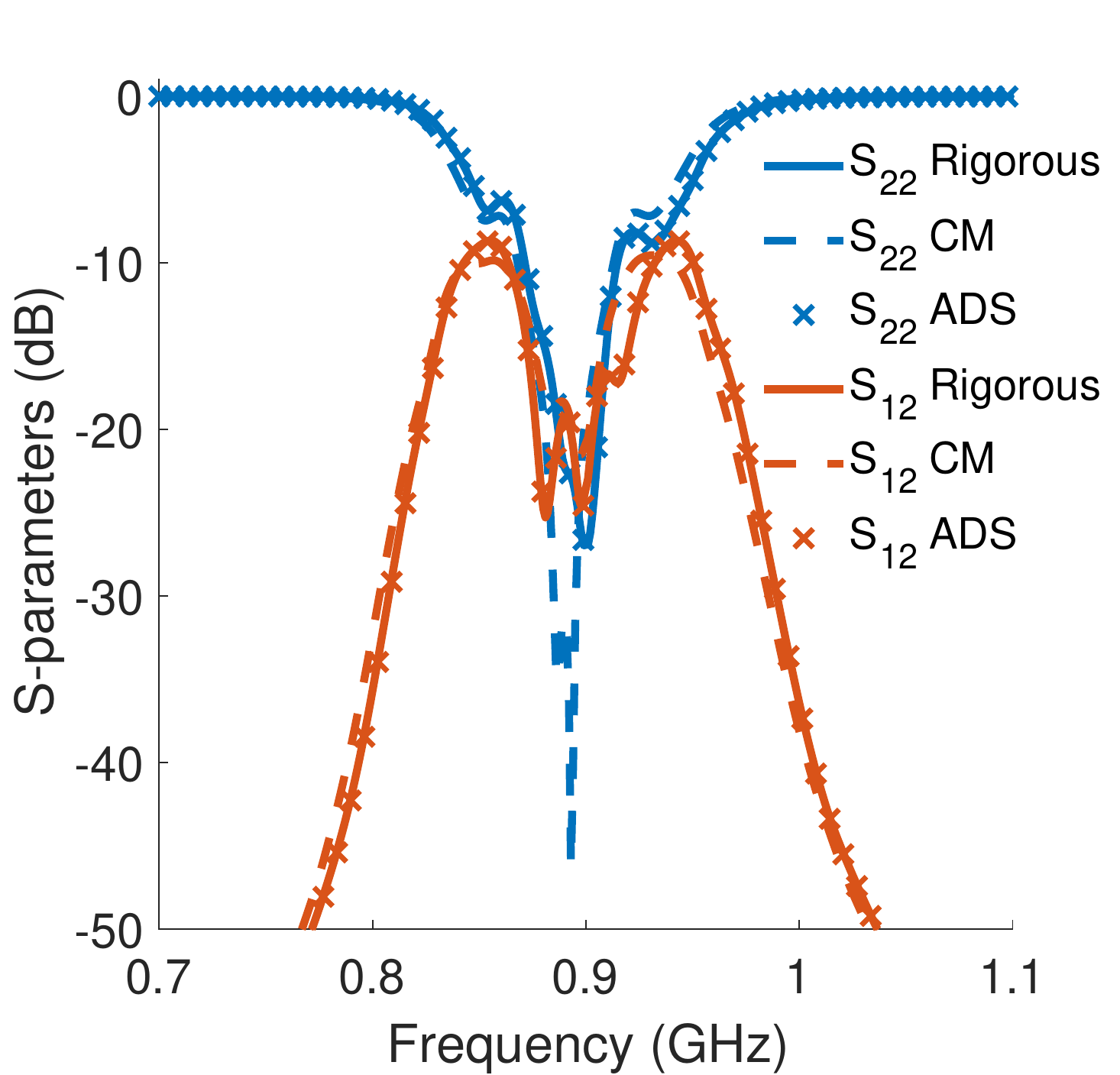}%
\label{fig:Simul_Order_4_Port2}
}
\caption{Scattering parameters of the
	fourth order non-reciprocal filter
	designed in Section~\ref{sec_num_results} computed using
	the commercial tool ADS (cross symbols),
	the coupling matrix
	approach introduced in this work with
	approximations
	(CM; dashed line)
	and without
	approximations	(solid lines).
	Approximations involve the use
	of~(\ref{eq:Admittance_approx})-(\ref{eq:fir_harmonic_resonators})
	and~(\ref{eq:inverter_non_reciprocal_lowpass}).
	In both calculations the number of harmonics has been fixed
	to $N_{har}=9$.}
\label{fig:Sparam_Rigorous_Comparison_Order4}
\end{figure}

As validation of the theory presented in this paper, we employ
this last filter design to compare our results with those
obtained with the commercial tool ADS \cite{ads2019}.
{\color{black}
Here we remark that the ADS results were obtained using
ideal built-in models to implement the time modulated
capacitors through~(\ref{eq:mod_capacitor}),
combined with the large signal scattering parameters analysis
module.
}
In addition, we also check what is the impact of the
approximations introduced in order to formulate the
frequency independent elements required by the
coupling matrix formalism. Essentially, the approximations
involve (i) the representation of the harmonic resonators with
the
{\color{black} frequency invariant susceptances}
of~(\ref{eq:fir_harmonic_resonators}),
instead of using the rigorous admittances given
in~(\ref{eq_yrk}); and (ii) the use of frequency independent
admittance inverters of~(\ref{eq:inverter_non_reciprocal_lowpass}),
instead of the rigorous expressions shown
in~(\ref{eq:inverter_non_reciprocal}).
Fig.~\ref{fig:Sparam_Rigorous_Comparison_Order4} compares the filter
response using these two different approaches, and using the
commercial tool ADS. It can be observed that our theory (Rigorous) agrees
perfectly with the results obtained with ADS. Small
differences can be observed between these two results
(ADS, Rigorous), and the results obtained introducing the approximations (CM). This
indicates that the impact of the approximations introduced
is indeed small, especially for narrowband filters.

\section{Practical Realization}
\label{sec_pract_realization}
\begin{figure}[!t]
\centering
\subfloat[]{\includegraphics[width=0.22\textwidth]{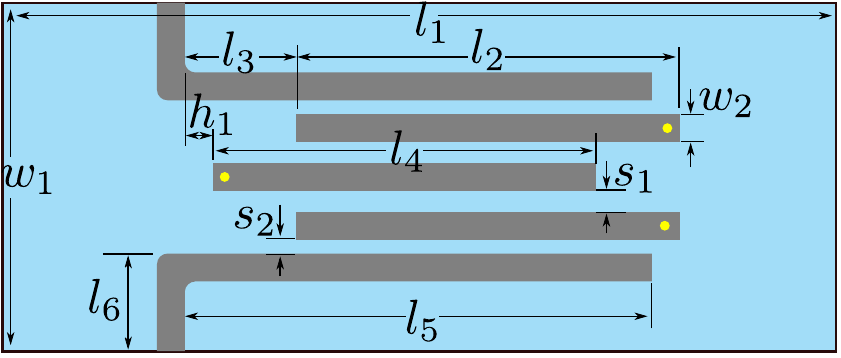}%
\label{fig:geom_third_order_top}
}
\hfill
\subfloat[]{\includegraphics[width=0.235\textwidth]{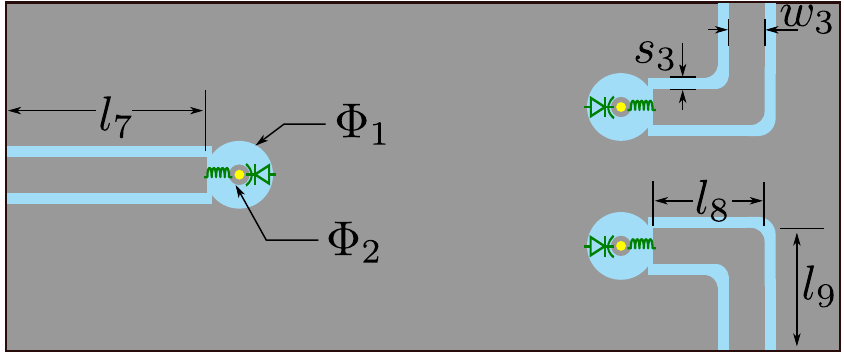}%
\label{fig:geom_third_order_bottom}
}
\hfill
\subfloat[]{\includegraphics[width=0.22\textwidth]{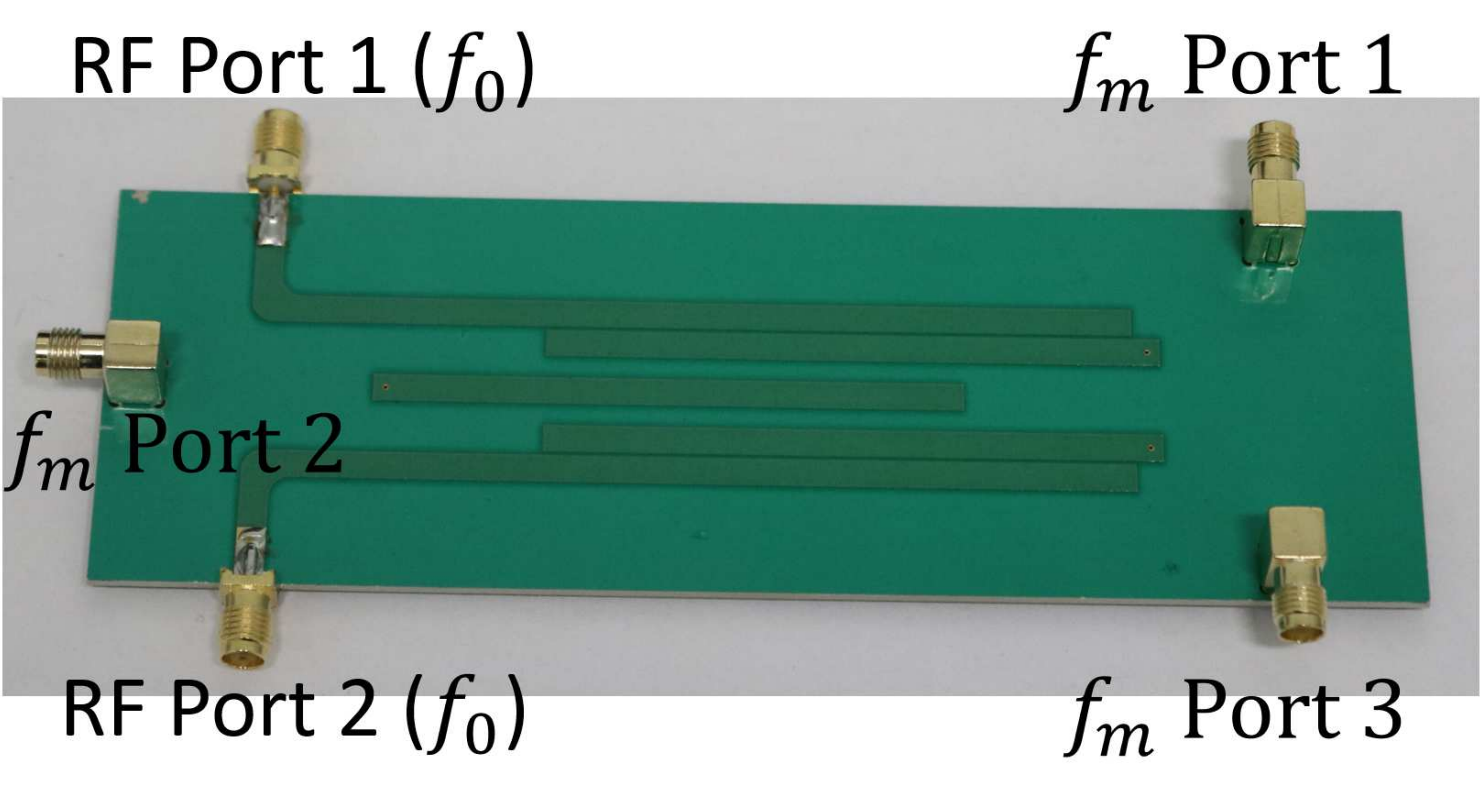}%
\label{fig:Filter3_top}
}
\hfill
\subfloat[]{\includegraphics[width=0.22\textwidth]{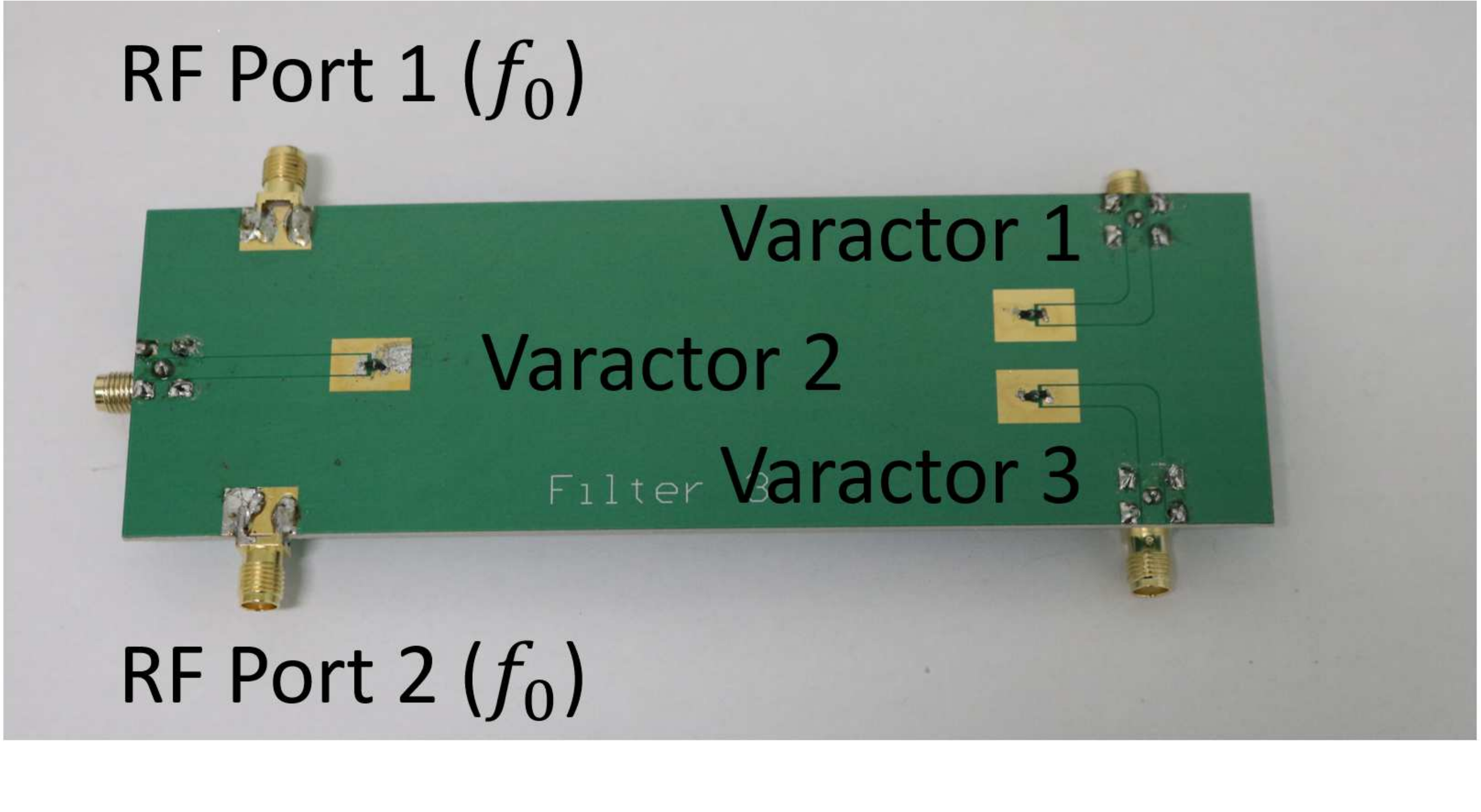}%
\label{fig:Filter3_bottom}
}
\caption{Geometry of the third order filter designed in microstrip
	technology. (a) Detail of the top metalization layer.
	(b) Detail of the
	bottom metalization layer.
	Panels (c)-(d) show a picture of the top and
	bottom metalization layers of the fabricated prototype, respectively.}
\label{fig:geom_third_order}
\end{figure}

\begin{figure}[!t]
\centering
\subfloat[]{\includegraphics[width=0.22\textwidth]
	{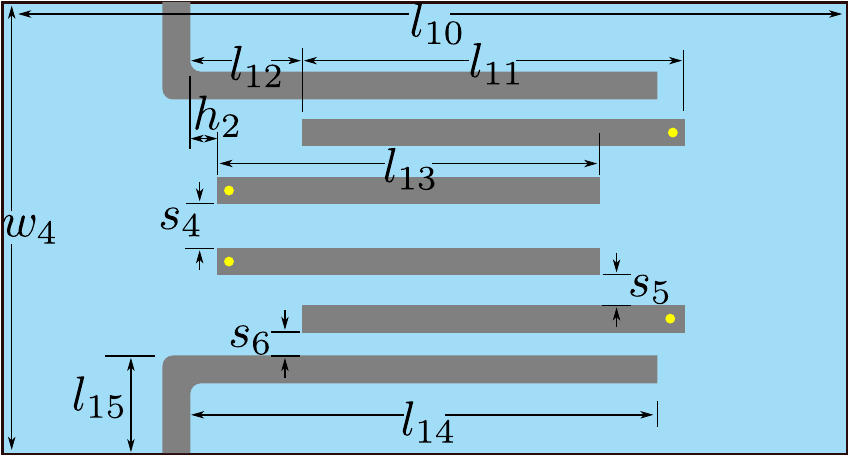}%
\label{fig:geom_fourth_order_top}
}
\hfill
\subfloat[]{\includegraphics[width=0.235\textwidth]
	{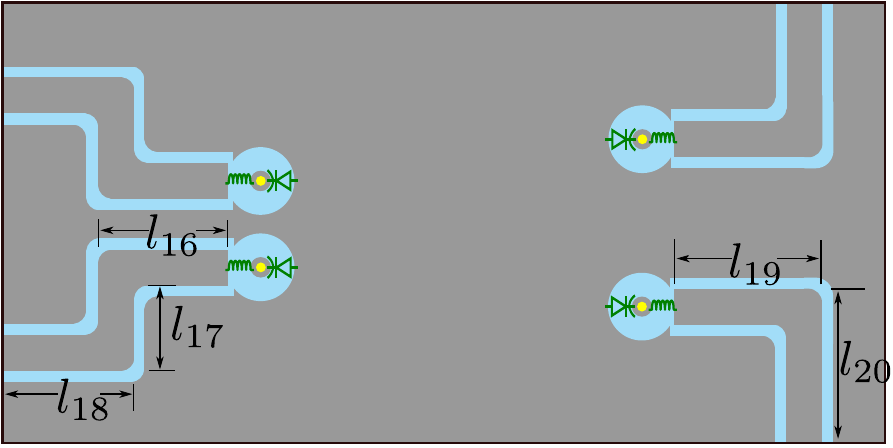}%
\label{fig:geom_fourth_order_bottom}
}
\hfill
\subfloat[]{\includegraphics[width=0.22\textwidth]{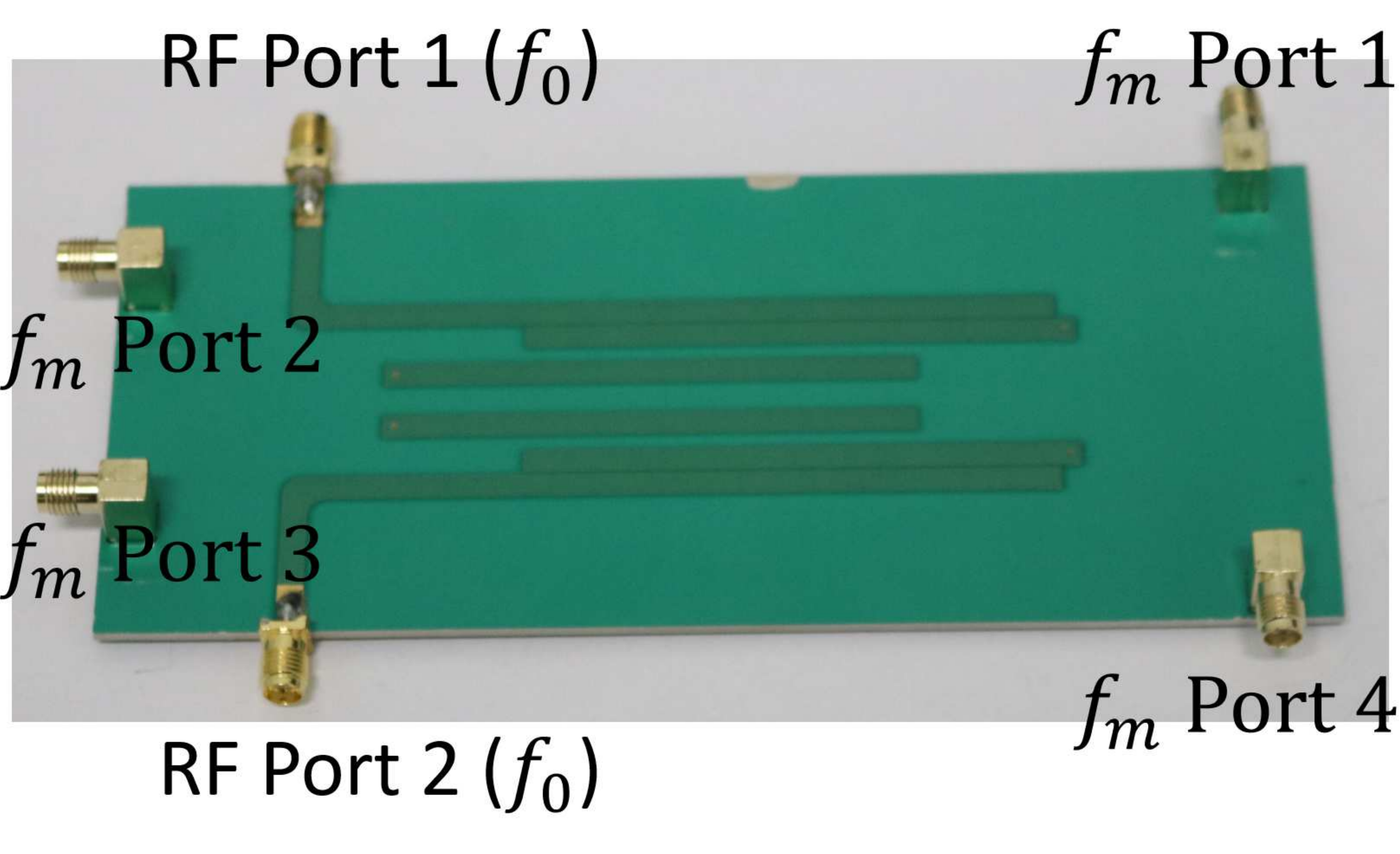}%
\label{fig:Filter4_top}
}
\hfill
\subfloat[]{\includegraphics[width=0.22\textwidth]{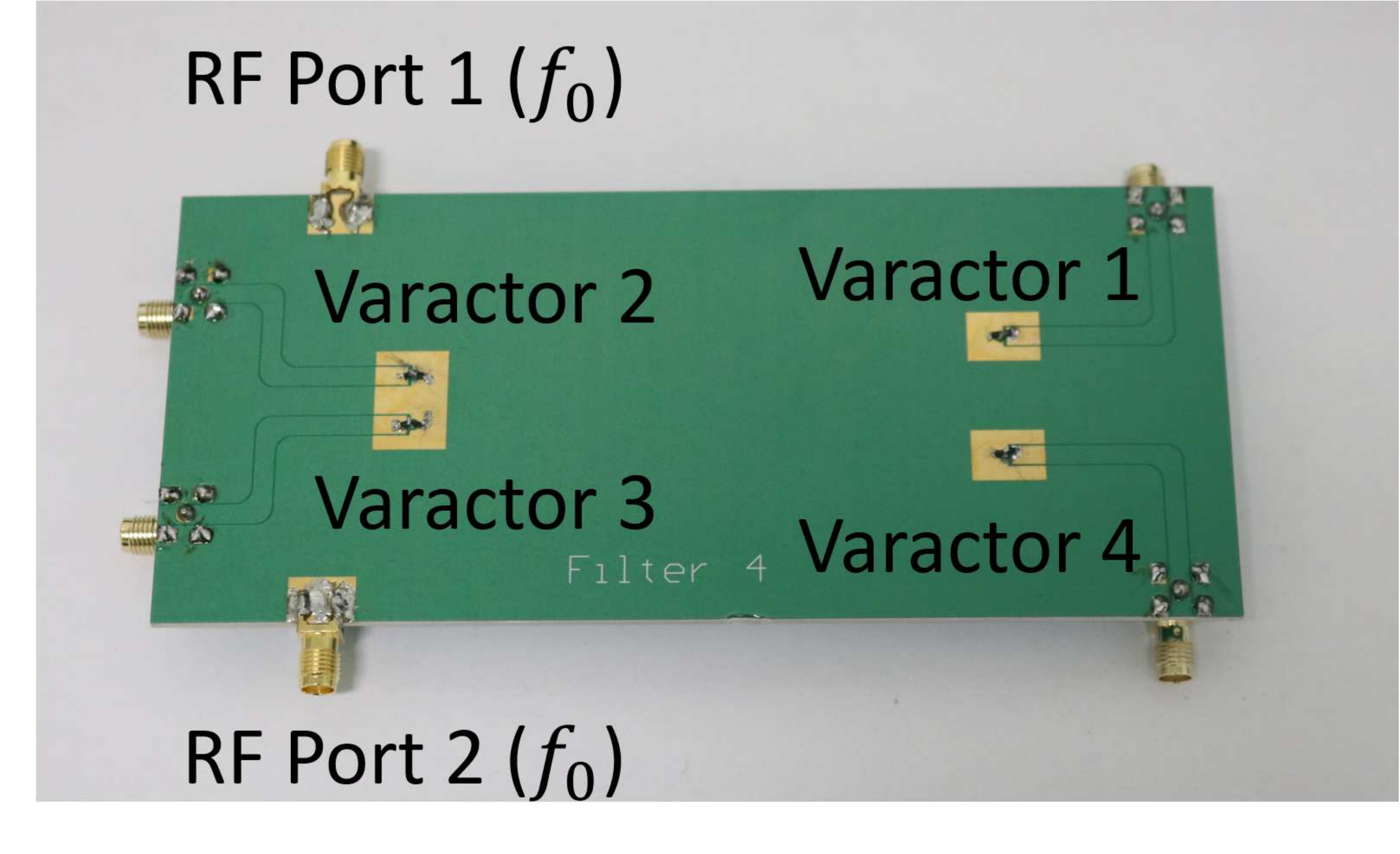}%
\label{fig:Filter4_bottom}
}
\hfill
\caption{Geometry of the fourth order filter designed in microstrip
	technology. (a) Detail of the top metalization layer.
	(b) Detail of the bottom
	metallization layer.
	Panels (c)-(d) show a picture of the top and bottom
	metalization layers of the fabricated prototype, respectively.}
\label{fig:geom_fourth_order}
\end{figure}

In this Section we present the fabrication and measurement of the
{\color{black}
two previously designed non-reciprocal filters,
}
implemented in microstrip technology. Fig.~\ref{fig:geom_third_order} and
Fig.~\ref{fig:geom_fourth_order} show the details of the filters together
with pictures of the manufactured prototypes. It can be observed
that the
{\color{black}
top metalization layer
}
contains the input/output RF feeding lines
and that the resonators are realized using quarter wavelength
transmission lines terminated on one side with a via-hole connected
to a varactor. On the
{\color{black}
bottom metalization layer,
}
the ground plane of the
microstrip line is modified to feed the various varactors
(from Skyworks, model SMV1234) with the corresponding modulating
signals using coplanar waveguides. In the figures we also show
the positions where the varactors are soldered in the board.
Note that a choke lumped inductor of value $180$~nH is
incorporated to increase the isolation between the signals
oscillating at $f_0$ and $f_m$. It should be emphasized that
this implementation enforces that the
RF and modulating signals
are supported on different planes of the substrates which
significantly increases the isolation between them
($>30$~dB). The substrate material used for the fabrication is
Rogers RT/duroid 6035 HTC with a relative dielectric constant $\epsilon_r=3.5$
and a thickness of $1.524$~mm. The final dimensions of the
fabricated prototypes are collected in Table~\ref{tab:dimensions_order3}
and Table~\ref{tab:dimensions_order4} for the third and fourth order
filters, respectively.
\begin{table}[!t]
\renewcommand{\arraystretch}{1.3}
\caption{Dimensions (in millimeters) of the fabricated
	3rd-order filter (see Fig.~\ref{fig:geom_third_order}).}
\label{tab:dimensions_order3}
\centering
\begin{tabular}{|c|c|c|c|c|c|c|c|c|} \hline
$W_1$ & $W_2$ & $W_3$ & $S_1$ & $S_2$ & $S_3$ & $h_1$ & $l_1$
	& $l_2$ \\ \hline
50   & 3.44  & 3  & 2.66 &  0.36 & 0.22 & 11.3 & 153 & 72  \\ \hline \hline
$l_3$&	$l_4$ & $l_5$ & $l_6$ & $l_7$ & $l_8$ & $l_9$ & $\Phi_1$
	& $\Phi_2$ \\ \hline \
31.3  & 69 & 100  & 17  &  31.95 & 15.5 & 20.4 & 1.8 & 1 \\ \hline
\end{tabular}
\end{table}
\begin{table}[!t]
\renewcommand{\arraystretch}{1.3}
\caption{Dimensions (in millimeters) of the fabricated
	4th-order filter (see Fig.~\ref{fig:geom_fourth_order}).}
\label{tab:dimensions_order4}
\centering
\begin{tabular}{|c|c|c|c|c|c|c|c|} \hline
$W_4$    & $S_4$     & $S_5$    & $S_6$    & $h_2$    & $l_{10}$
         & $l_{11}$  & $l_{12}$  \\ \hline
70       & 4.56      & 2.21     & 0.21     & 9.3      & 160
         & 73        & 27.8 \\ \hline \hline
$l_{13}$ & $l_{14}$  & $l_{15}$ & $l_{16}$ & $l_{17}$ & $l_{18}$
	 & $l_{19}$  & $l_{20}$ \\ \hline
70.3     & 98.1      & 23.4     & 22.5     & 14       & 17
	 & 25.5      & 26.9  \\ \hline
\end{tabular}
\end{table}

\begin{figure*}[!t]
\centering
\subfloat[]{\includegraphics[width=0.325\textwidth]
	{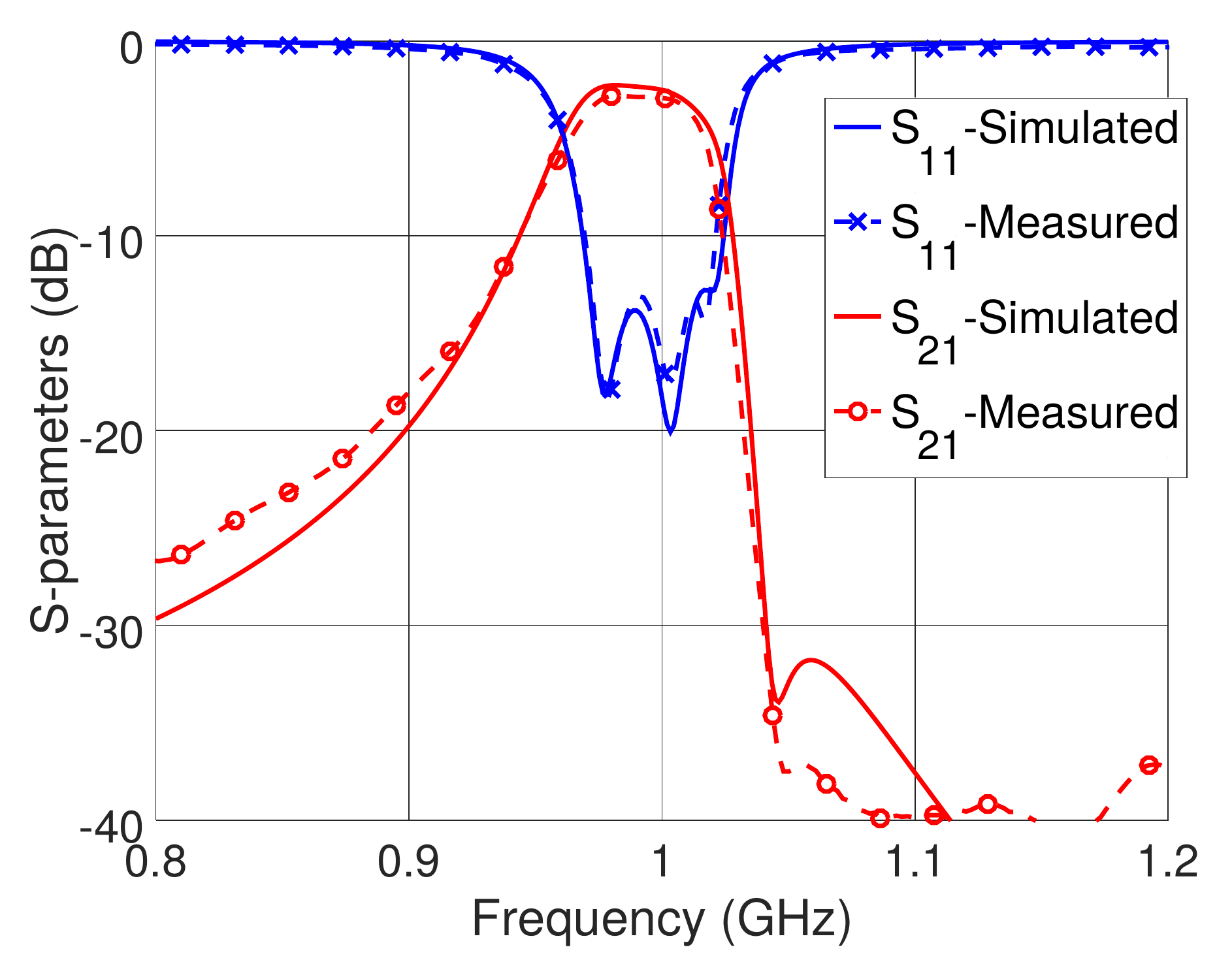}
\label{fig:Measured_Order_3_nonmodulated}
}
\hfill
\subfloat[]{\includegraphics[width=0.325\textwidth]
	{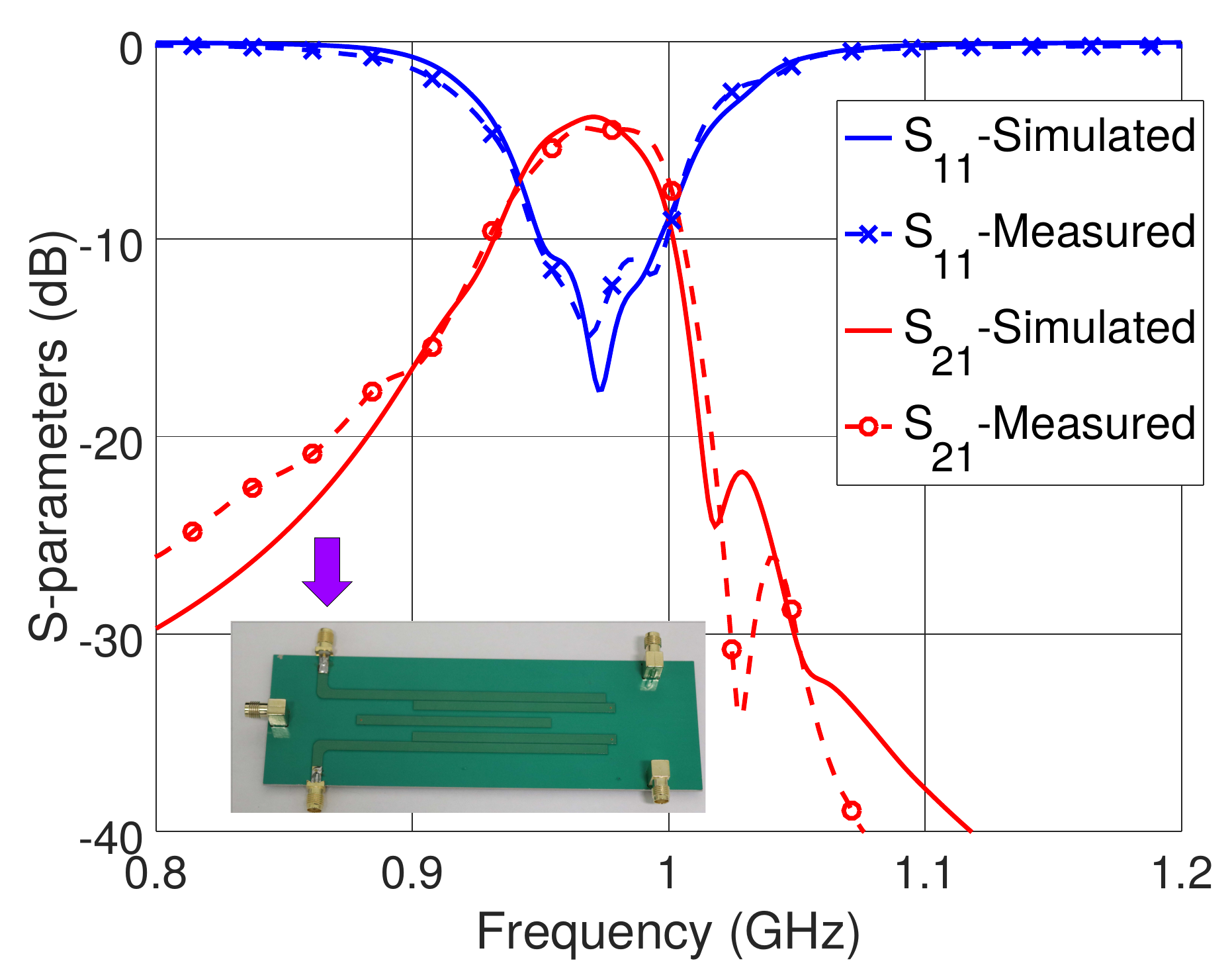}
\label{fig:Measured_Order_3_modulated_port1}
}
\hfill
\subfloat[]{\includegraphics[width=0.325\textwidth]
	{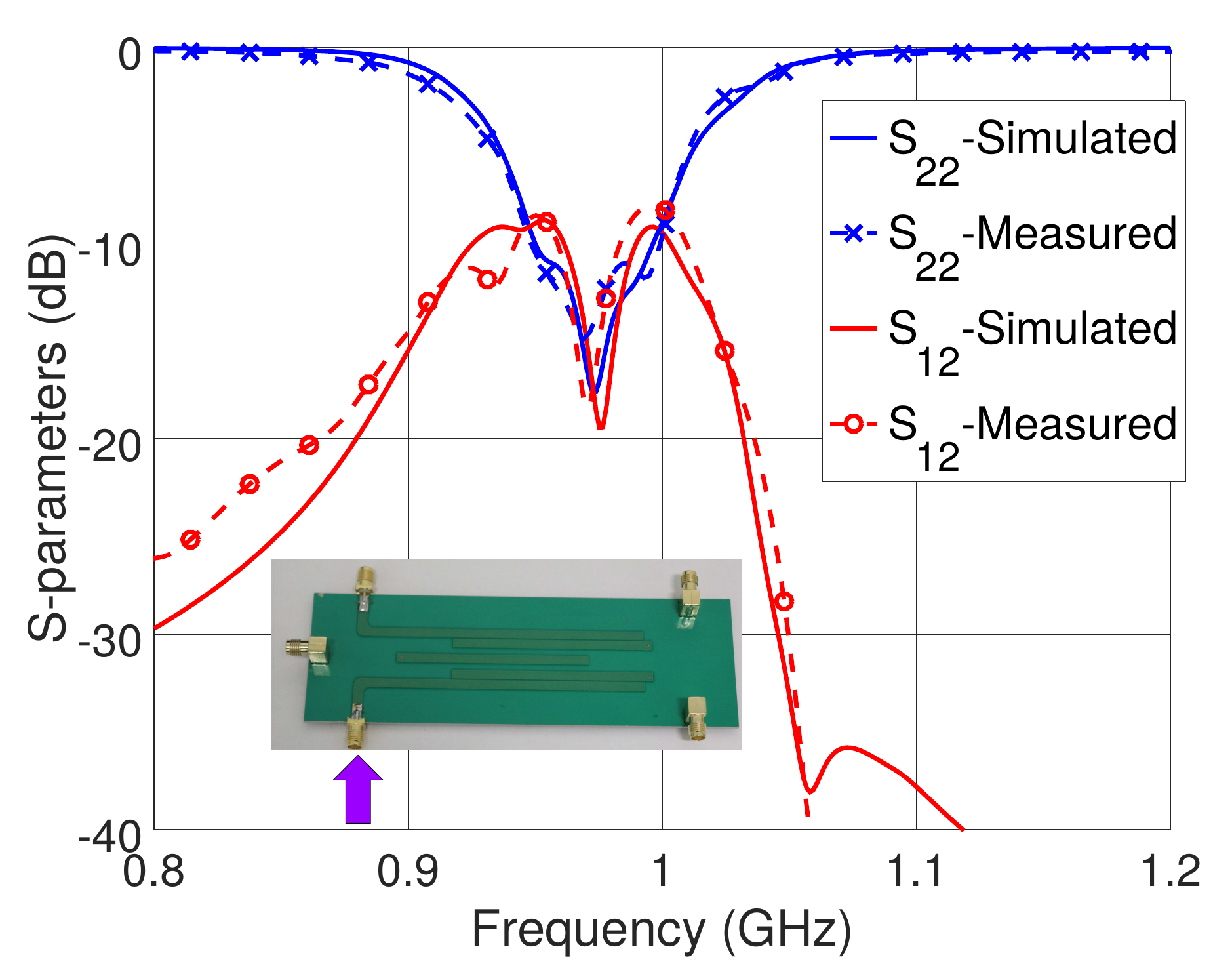}
\label{fig:Measured_Order_3_modulated_port2}
}
\hfill
\caption{Measured response of the manufactured third order non-reciprocal
	filter and comparison with respect to the numerical results
	obtained with the proposed technique
	{\color{black}
	(losses and parasitic cross couplings have been included
	in the coupling matrix approach).}
	(a) Unmodulated response
	{\color{black}
	(note that in this case $S_{11}=S_{22}$ and $S_{12}=S_{21}$).
	}
	(b)-(c) Response obtained when the modulating signal is
	applied to the varactors and the filter is excited
	(shown in the inset using a magenta arrow) from the
	first (b) and the second (c) port.}
\label{fig:Measured_Order_3}
\end{figure*}

Fig.~\ref{fig:Measured_Order_3_nonmodulated} shows the measured
results for the third order filter in the absence of any modulation
and compares them with the
simulated response using the coupling matrix formalism.
{\color{black}
Here we should remark that the simulated responses of the
filters are all obtained with the theoretical analysis
presented in Section~\ref{sec_network_time_mod_res}.
}
{\color{black}
In addition, it can be observed in the measured response
some deviations with respect to the response of the designed prototype
shown in Fig.~\ref{fig:Sparam_Simul_Order_3}. The differences
are mainly due to the insertion losses within the passband,
which are around $IL=2.8$~dB, and to some parasitic cross couplings
that were not taken into account during the initial design.
These two factors have been included in the simulated responses
obtained with the coupling matrix formalism derived in this work,
shown in Fig.~\ref{fig:Measured_Order_3}. Losses in the resonators
are modeled with an additional resistor connected in parallel.
The response show in Fig.~\ref{fig:Measured_Order_3_nonmodulated} is used
to extract the unloaded quality factors of the resonators, giving
$Q_U=114$. This unloaded quality factor is small, but it is not
uncommon of planar technology \cite{yan18}, and especially
when using microstrip line printed resonators. In addition, we have
found that the drop of selectivity in the lower side of the passband
is mainly due to a non negligible cross coupling between the
ports and the second resonator, giving normalized coupling
factors or $M_{P_1 2}=M_{2 P_2}=0.26$. Although of much weaker value,
there is also a small parasitic coupling between the first and
third resonator, which is modeled with a normalized coupling factor
of $M_{13}=0.09$. It can be observed that the agreement between
measured and simulated results are reasonably good, once losses
and parasitic couplings are included in the derived coupling
matrix formalism.
}

\begin{figure*}[!t]
\centering
\subfloat[]{\includegraphics[width=0.325\textwidth]
        {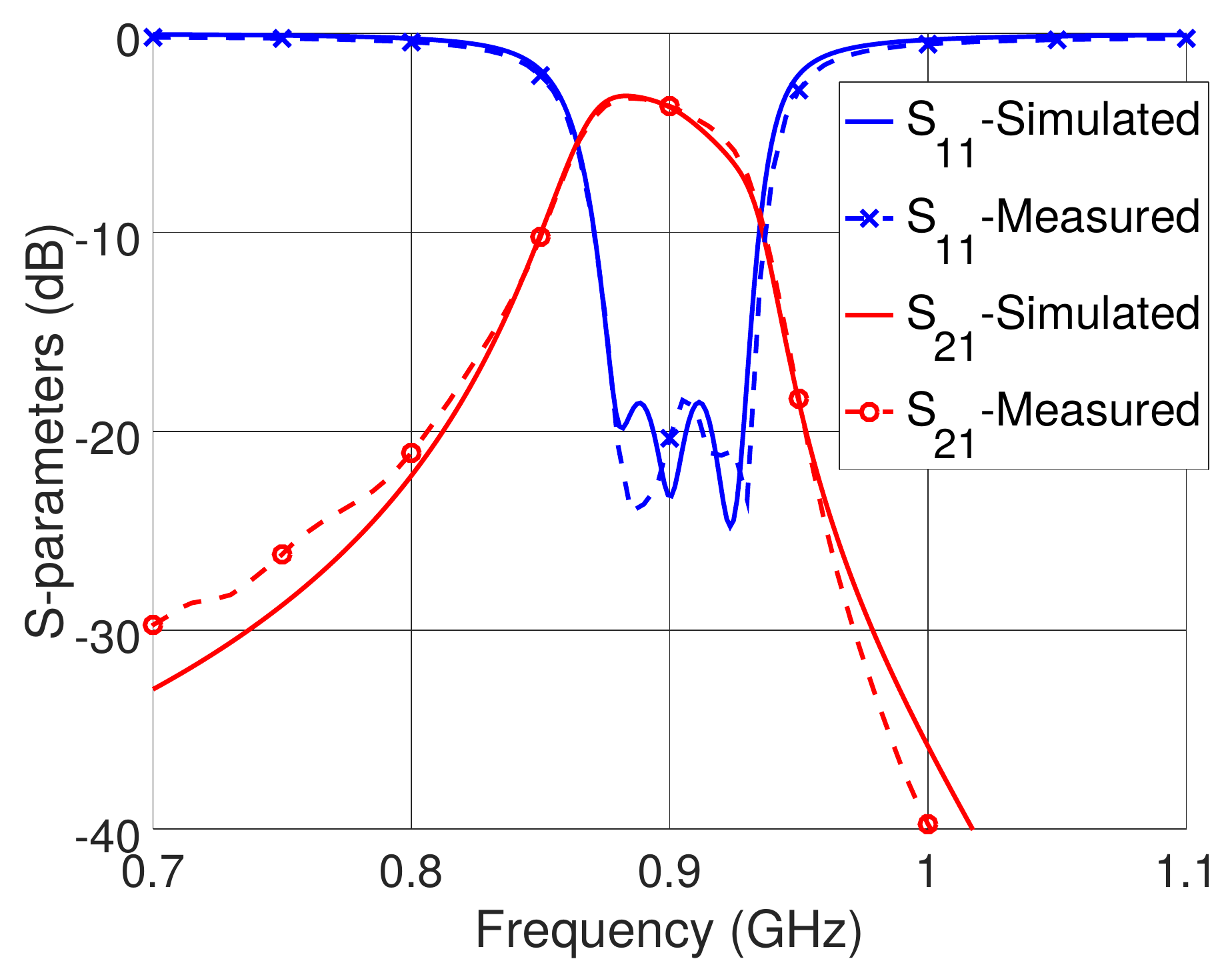}
\label{fig:Measured_Order_4_nonmodulated}
}
\hfill
\subfloat[]{\includegraphics[width=0.325\textwidth]
        {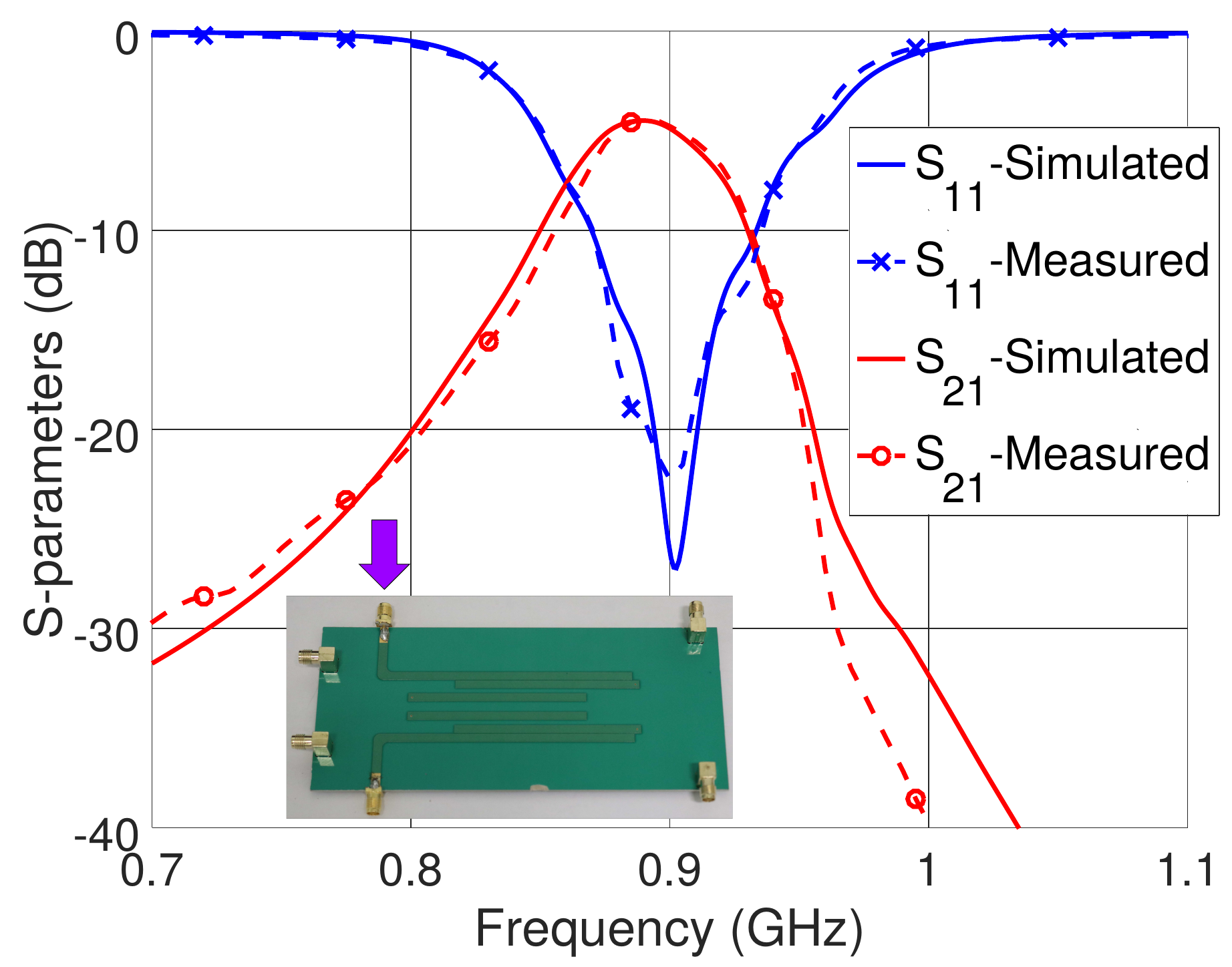}
\label{fig:Measured_Order_4_modulated_port1}
}
\hfill
\subfloat[]{\includegraphics[width=0.325\textwidth]
        {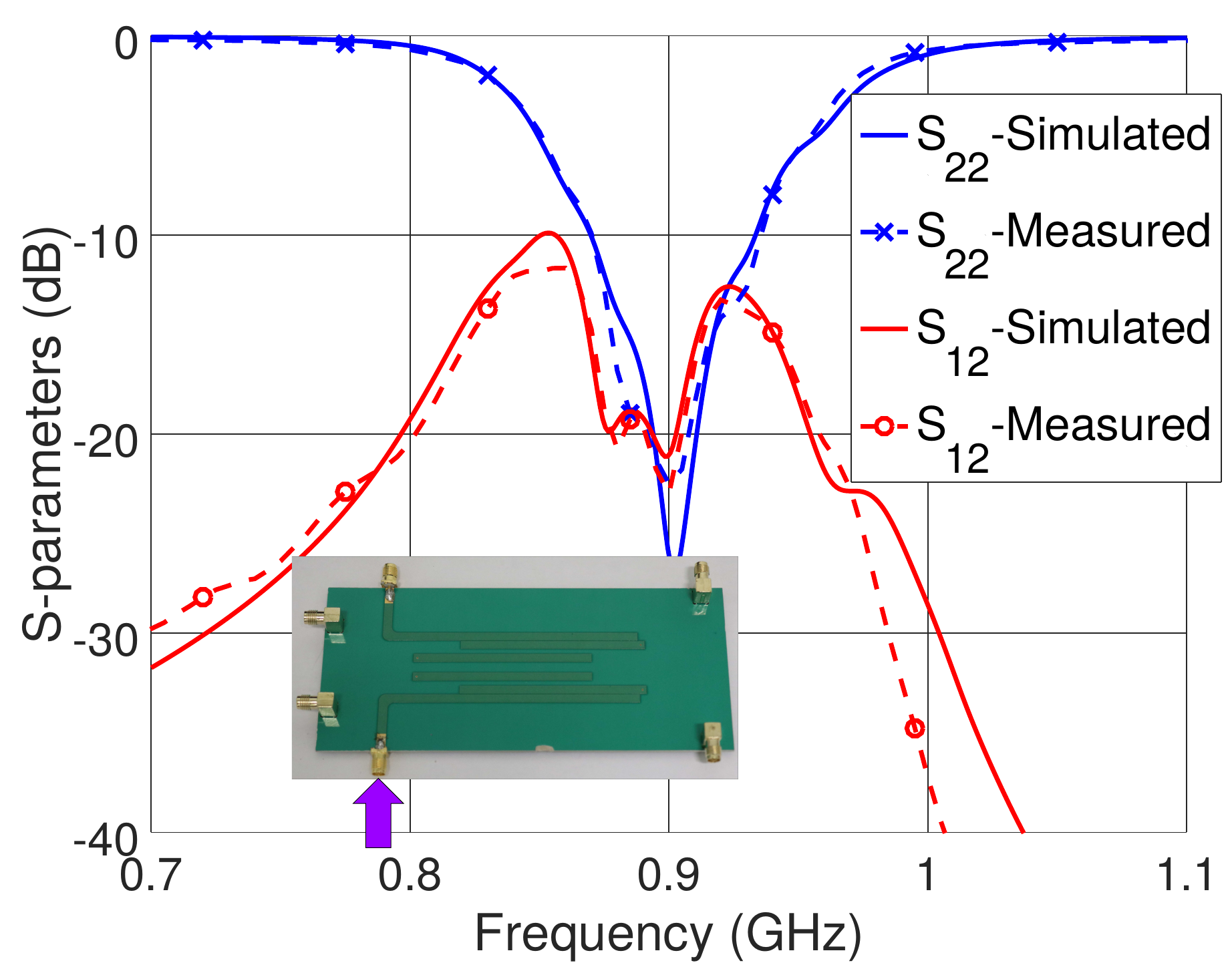}
\label{fig:Measured_Order_4_modulated_port2}
}
\caption{Measured response of the manufactured fourth order
	non-reciprocal filter and comparison with respect to the
	numerical results obtained with the proposed technique
	{\color{black}
	(losses and parasitic cross couplings have been included
	in the coupling matrix approach).}
	(a) Unmodulated response
	{\color{black}
        (note that in this case $S_{11}=S_{22}$ and $S_{12}=S_{21}$).
        }
	(b)-(c) Response obtained when
	the modulating signal is applied to the varactors and the
	filter is excited (shown in the inset using a magenta arrow)
	from the first (b) and the second (c) port.}
\label{fig:Measured_Order_4}
\end{figure*}

Fig.~\ref{fig:Measured_Order_3_modulated_port1} presents the measured
versus simulated results when the modulating signal is
{\color{black}
applied to
}
the varactors and the filter is excited from the first port.
It can be observed that the filter behaves as in the unmodulated case,
with increased losses of around $IL=4.5$~dB that account for both
dissipation effects and the power converted into nonlinear harmonics.
The useful bandwidth measured at a return loss level
of $RL=11$~dB is $45$~MHz.
Fig.~\ref{fig:Measured_Order_3_modulated_port2} shows the response
of the prototype when it is excited from the second port.
The filtering response is suppressed and instead the device
behaves as an isolator that attenuates all incoming power.
Maximum non-reciprocity is achieved at the center of
the passband with a directivity of $D_0=13.8$~dB.
{\color{black}
It can be observed that
when losses and parasitic cross couplings are included in the
coupling matrix model,
a good agreement
is maintained between
measured data and numerical simulations.
}

Measurements corresponding to the fourth order filter are shown
in Fig.~\ref{fig:Measured_Order_4}.
Fig.~\ref{fig:Measured_Order_4_nonmodulated} plots the response of
the filter before introducing the modulating signal and compares
it with respect to the response of the ideal circuit. Again
the bandwidth and the ripple level obtained within the passband
are very similar. Measured results exhibit a perfectly constant
equi-ripple response, since the resonant frequencies of the
resonators are slightly tuned with constant voltages applied to
the varactors. The insertion losses due to dissipation effects
in the resonators and in the varactors are slightly larger than
in the previous filter, obtaining a minimum level of $IL=3.2$~dB
that slowly increases towards the end of the passband.
{\color{black}
The insertion losses measured in the unmodulated case
(Fig.~\ref{fig:Measured_Order_4_nonmodulated}) were used
again to extract the unloaded quality factor of the
resonators, obtaining essentially the same value
as in the previous example. This is something to be
expected, since the same resonators as before were used
in this second prototype, and the same technology
was used for manufacturing. In any case, this also
shows high repeatability of the employed manufacturing process.

Measured results again show a drop in selectivity as compared
to the designed response of Fig.~\ref{fig:Sparam_Simul_Order_4},
especially in the lower side of the passband. Once more we found
that this is due to parasitic cross couplings not taken into
account during the initial design process. In the comparison
shown in Fig.~\ref{fig:Measured_Order_4}, we can observe
good agreement between measured and simulated responses
when losses and parasitic couplings are included in the
derived model. Again, we found that the strongest parasitic
couplings occur between the ports and the closest non
contiguous resonators: $M_{P_1 2}=M_{3 P_2}=0.23$ and
$M_{P_1 3}=M_{2 P_2}=0.1$.
However, non negligible parasitic
couplings have also been found between internal resonators:
$M_{13}=M_{24}=0.12$ and $M_{14}=0.06$.
}
\begin{table}[!t]
\renewcommand{\arraystretch}{1.3}
\caption{Basic electrical performances obtained for the
	two manufactured filters.}
\label{tab:summary}
\centering
\begin{tabular}{|c|c|c|c|c|} \hline
	     & $IL$~(dB)& $RL$~(dB)& $D$~(dB)& $F_B$~(\%) \\ \hline \hline
Third order  & 4.5      & 11       & 13.8    &  4.6     \\ \hline
Fourth order & 4.4      & 11       & 13.6    &  6.4     \\ \hline
\end{tabular}
\end{table}

Fig.~\ref{fig:Measured_Order_4_modulated_port1} presents the
measured results obtained from the manufactured prototype when
the modulating signal is
{\color{black}
applied to
}
the varactors and the filter
is excited from the first port.
The fabricated prototype behaves as a filter with a useful
bandwidth of $57$~MHz measured at a return loss level of $RL=11$~dB.
With respect to the unmodulated case, the insertion losses in the
forward direction have increased to $IL=4.4$~dB. As in the previous case,
the extra losses are due to power converted into nonlinear harmonics
that is not converted back to the fundamental frequency. Exciting
the device from the second port significantly attenuates the
propagating energy. The strong non-reciprocity predicted by
the initial simulations is confirmed by the measurements. Around
the center frequency of the passband, the directivity is better
than $D_0=13.6$~dB in a bandwidth of $35$~MHz. Across the entire
passband, the directivity is always better than $D=7.2$~dB.
{\color{black}
In general, good agreement between measured and
simulated responses are obtained when losses and parasitic
couplings are included in the derived coupling matrix model.
}
{\color{black}
For reference, the basic performances for both manufactured filters
are collected in Table~\ref{tab:summary}.
}

{\color{black}
Another important characteristic of these devices for many
applications is the power handling levels. The hardware built
could not be tested under high power signals. Primarily,
the power handling will be limited by the technology used
to build a similar unmodulated circuit  \cite{wu09}.
However, an interesting future research
topic will be the assessment on
how the additional circuitry needed by modulation
signals and the presence of varactors
affect the power handling levels, and which
arrangements are more appropriate to reduce these undesired effects.
}
\section{Conclusion}
We have presented the analysis of non-reciprocal filters based
on time modulated capacitors using a coupling matrix formalism.
From the initial topology of the filter, a novel coupling topology
using harmonic resonators is first derived. Closed form analytic
expressions have been obtained to represent the harmonic resonators
with
{\color{black} frequency invariant susceptances},
thus obtaining the diagonal elements of the
traditional coupling matrix. Also, non-reciprocal admittance
inverters have been analytically computed to account for the
couplings between harmonic resonators, thus obtaining the
off-diagonal elements of the coupling matrix. The derived analysis
method has been validated with the design and fabrication of third
and fourth order filters implemented in microstrip technology.
Measured results on the fabricated prototypes, and results
obtained with a commercial tool are found to agree
well with respect to numerical calculations obtained using the new coupling
matrix formulation, thus fully
validating the theory presented.
\section*{Acknowledgment}
Authors are grateful to Rogers Corporation for the generous donation
of the dielectrics employed in this work.
\ifCLASSOPTIONcaptionsoff
  \newpage
\fi

\end{document}